\documentclass[%
 reprint,
 amsmath,amssymb,
 aps,
 superscriptaddress,
]{revtex4-2}

\usepackage{graphicx}
\usepackage{dcolumn}
\usepackage{bm}
\usepackage{subfig}
\usepackage{booktabs}
\usepackage{multirow}
\usepackage{float}
\usepackage[linesnumbered,ruled,vlined]{algorithm2e}
\usepackage{algpseudocode}
\usepackage{mathtools}

\newcommand{\IQSE}{Shenzhen Institute for Quantum Science and Engineering, Southern University of Science and Technology, Shenzhen 518055, China}
\newcommand{\IQA}{International Quantum Academy, Shenzhen 518048, China}
\newcommand{\GKEY}{Guangdong Provincial Key Laboratory of Quantum Science and Engineering, Southern University of Science and Technology, Shenzhen 518055, China}
\newcommand{\SKEY}{Shenzhen Key Laboratory of Quantum Science and Engineering, Southern University of Science and Technology, Shenzhen 518055, China}

\newcommand{\FD}{Department of Physics, Fudan University, Shanghai, 200433, China}
\newcommand{\THU}{Center for Intelligent and Networked Systems, Department of Automation, Tsinghua University, Beijing, 100084, China.}

\newcommand{\BUPT}{State Key Laboratory of Networking and Switching Technology, Beijing University of Posts and Telecommunications, Beijing, 100876, China}

\newcommand{\ZJU}{Key Laboratory of Micro-Nano Electronic Devices and Smart Systems of Zhejiang Province, College of Information Science and Electronic Engineering, Zhejiang University, Hangzhou 310027, Zhejiang, China}

\begin{document}

\preprint{APS/123-QED}

\title{High-Order Modulation Large MIMO Detector Based on Physics-Inspired Methods}
\thanks{A footnote to the article title}%

\author{Qing-Guo Zeng}
\affiliation{\IQSE}

\author{Xiao-Peng Cui}
\affiliation{\FD}

\author{Xian-Zhe Tao}
\affiliation{\IQSE}

\author{Jia-Qi Hu}
\affiliation{\THU}

\author{Shi-Jie Pan}
\affiliation{\BUPT}

\author{Wei E. I. Sha}
\email{ weisha@zju.edu.cn}
\affiliation{\ZJU}

\author{Man-Hong Yung}
\email{ yung@sustech.edu.cn}
\affiliation{\IQSE}
\affiliation{\IQA}
\affiliation{\GKEY}
\affiliation{\SKEY}



\date{\today}

\begin{abstract}
Applying quantum annealing or current quantum-/physics-inspired algorithms for MIMO detection always abandon the direct gray-coded bit-to-symbol mapping in order to obtain Ising form, leading to inconsistency errors. This often results in slow convergence rates and error floor, particularly with high-order modulations. We propose HOPbit, a novel MIMO detector designed to address this issue by transforming the MIMO detection problem into a higher-order unconstrained binary optimization (HUBO) problem  while maintaining gray-coded bit-to-symbol mapping. The method then employs the simulated probabilistic bits (p-bits) algorithm to directly solve HUBO without degradation. This innovative strategy enables HOPbit to achieve rapid convergence and attain near-optimal maximum-likelihood performance in most scenarios, even those involving high-order modulations. The experiments show that HOPbit surpasses ParaMax by several orders of magnitude in terms of bit error rate (BER) in the context of 12-user massive and large MIMO systems even with computing resources. In addition, HOPbit achieves lower BER rates compared to other traditional detectors.
\end{abstract}

\maketitle


\section{\label{sec:intro}Introduction}

The past decade has witnessed an exponential growth of global mobile data traffic along with the tremendous rise in the number of mobile users and data-intensive content~\cite{albreem2019massive}. This urges the fast development of wireless technologies. Multiple-input multiple-output (MIMO) is a key technique to satisfy the massive demand for wireless data traffic in current and future generation of wireless systems in which multiple antennas at both transmitter and receiver are combined to improve the capacity of a radio link based on multipath propagation. 

However, it is challenging for the receiver to reproduce the transmitted symbols based on channel state information (CSI) and the received signal with the presence of noise and interference, which is referred to as MIMO detection problem~\cite{trotobas2020review}. The exact algorithm, Maximum Likelihood Detector (MLD), has been proven to be NP-hard~\cite{yang2015fifty}. Linear detectors such as Zero-Forcing (ZF) and Minimum Mean Squared Error (MMSE) perform nearly to MLD in massive MIMO systems at the cost of a large number of receiver antennas to support relatively fewer users~\cite{larsson2014massive}. That is, their detection accuracy becomes poor in large MIMO systems where the number of receiver antennas is the same as that of user antennas~\cite{nikitopoulos2014geosphere}. Despite decades of research, practical methods with near-optimal detection accuracy for larger MIMO systems remain elusive~\cite{trotobas2020review}.

In recent years, some researchers have begun to investigate the application of quantum annealing to MIMO detection~\cite{kim2021heuristic}. QuAMax, the first quantum annealing-based MIMO detector uses a linear transformation function to reduce the MIMO detection problem into Ising problem and then exploits quantum annealer to solve it~\cite{kim2019leveraging}. Tabi et al. improve QuAMax by combining Single Qubit Correction post-processing technique and apply it to larger MIMO systems and higher-order modulation (64-QAM) cases \cite{tabi2021evaluation}. As a result, Correction-enhanced QuAMax detector achieves a lower bit error rate (BER). Quantum Annealing-based method is promising and has the potential to achieve polynomial or exponential speedups with extremely near-optimal performance in large MIMO systems~\cite{djidjev2018efficient, crosson2016simulated, albash2018demonstration}. Considering the sensitivity of quantum devices to noise, a hybrid quantum-classical detector has been proposed in~\cite{kim2020towards} to combine the advantages of both hardware. However, in Noisy Intermediate-Scale Quantum (NISQ) era, these are not optimistic for practical applications due to the limited qubit connectivity and physical noises~\cite{hegade2021shortcuts}. Fortunately, some Ising machines and quantum/physical-inspired algorithms such as Noisy Mean-Field Algorithm~\cite{king2018emulating}, Coherent Ising machine and its simulated algorithms ~\cite{marandi2014network, tiunov2019annealing,reifenstein2021coherent}, Simulated Bifurcation Machine~\cite{ goto2021high}, et., have demonstrated their superiority over the current quantum computing devices in specific problems~\cite{mcmahon2016fully}. By adopting the same transformation, these methods/solvers can also solve MIMO detection problem. The detector, ParaMax employs the parallel tempering algorithm to solve Ising problems. One key advantage of this approach is its capacity to generate soft output based on confidence information that is useful for iterative MIMO detection processes~\cite{10.1145/3447993.3448619}. Thus, the author~\cite{10.1145/3447993.3448619} also proposes 2R-ParaMax, a MIMO detector that uses parallel tempering algorithm in twice rounds. It fixes some spins based on spinwise detection confidence and updates the rest spins in the second round. Sing et al. begin to investigate the disadvantage of straightforward ML-to-Ising transformation and find out an error floor problem~\cite{singh2022ising}. As a result, they add a regularized term to mitigate this issue. In order to reduce MLD to Ising form, all of the aforementioned methods have to give up 2-dimensional gray code mapping, resulting in the inconsistency of change between the objective function and the hamming distance of bit strings. We refer to the error due to this inconsistency as the inconsistency error. The higher-order modulation, the larger the inconsistency error is, thereby reducing the detection accuracy. In addition, Norimoto et al. point out that the methods based on the linear transformation function need extra encoding and decoding steps to transform the solution back to Gray-coded form~\cite{norimoto2023quantum}. As a result, they directly use gray-coded bit-to-symbol mapping to reduce MIMO detection problem into high-order unconstrained binary problem (HUBO), and then solve it by utilizing a quantum algorithm, grover adaptive search~\cite{gilliam2021grover}. However, Grover's algorithm remains more of a theoretical milestone than a practical tool since its meaningful acceleration requires the use of a fault-tolerant quantum computer with a large number of qubits , which are still far away~\cite{stoudenmire2023grover}.

Hence, we propose HOPbit, a detector that also transforms MIMO detection problem to high-order unconstrained binary problem (HUBO) while maintaining the gray-coded bit-to-symbol mapping to mitigate error floor. The current quantum annealing or quantum/physics-inspired methods for HUBO usually introduce auxiliary variables and transform it into QUBO, which suffers from the proportional increases in auxiliary bits and are extra terms with the number of antennas~\cite{rodriguez2018linear, mandal2020compressed,verma2022penalty,ayodele2022penalty,garcia2022exact}. In this work, we implement a simulator of high-order p-bits to solve it directly. Besides, we apply precomputation of the input signal of p-bit for further acceleration.

The rest of this article is organized as follows. MIMO detection problem is described in the next section. In Section 3, we presented each component and detailed process of the HOPbit. Its performances are investigated and compared with the QUBO solver and other conventional detectors in Section 4. Finally, we conclude and discuss the future work in the last section.

\section{\label{sec:mimo}MIMO detection problem}

We assume that in MIMO detection, $N_t$ users each with single antenna transmit the symbols $\bm{x}=[x_1,x_2,\dots, x_{N_t}]^\top\in\mathbb{C}^{N_t}$ from constellation $\Omega$ and then the symbols is received by base station with $N_r$ antennas. The received symbols can be presented $\bm{y}=[y_1, y_2,\dots, y_{N_r}]^\top\in\mathbb{C}^{N_r}=\bm{y}^R+j\bm{y}^I$. The transmission characteristic between $N_t$ user antennas and $N_r$ receiver antennas is summarized in a channel matrix $\bm{H}\in\mathbb{C}^{N_r\times N_t}=\bm{H}^R+j\bm{H}^I$. Finally, this process can be formulated as 
\begin{equation}
    \bm{y}=\bm{H}\bm{x}+\bm{n},
\end{equation}
where $\bm{n}\in\mathbb{C}^{N_r}$ indicates the additive white Gaussian noise. The MIMO detection problem is to reproduce the transmission symbols when given the channel matrix and received symbols as far as possible.

With the presence of additive noise, the ML detector is to seek the optimal solution that minimizes the detection error as followed
\begin{equation}
    \hat{\bm{x}}_{\text{ML}}=\arg\min_{\bm{x}\in\Omega^{N_t}}||\bm{y}-\bm{H}\bm{x}||^2.\label{equ:ml}
\end{equation}
It means that the detector searches for the solution in an exponentially large space with the size of the constellation and the number of user antennas even equipped with sphere decoder reduction in search space~\cite{damen2003maximum}.

\section{HOPbit}
It is impractical to apply ML detector for massive and large MIMO systems. In this section, we propose a novel detector, HOPbit that reduces ML detection to HUBO problems, thereby eliminating the inconsistency error. Firstly, a standard HUBO problem form and the transformation of ML into HUBO problem are given. Then, we introduce a simulated p-bits solver. Finally, we describe the complete pipeline of HOPbit.

\subsection{ML-to-HUBO transformation}
A HUBO problem is to find the minimization of a pseudo-Boolean energy function (Hamiltonian)~\cite{hammer1963determination, boros2002pseudo}:
\begin{align}
H(\bm{q})=&\sum_{i_1}J_{i_1}q_{i_1}+\sum_{i_1<i_2}J_{i_1i_2}q_{i_1}q_{i_2}+\cdots\notag\\
&+\sum_{i_1<i_2<\cdots<i_k}J_{i_1i_2\cdots i_k}\prod_{1}^kq_{j},
    \label{equ:H}
\end{align}
where $N$ is the number of binary variables, $k$ is the function degree, $\bm{q}=[q_1,\dots, q_N]\in\{-1, 1\}^N$ denotes the vector of all the variables and the coefficients $J_{i_1\cdots i_k}$ are real numbers. 

The most critical thing for reducing the ML problem into HUBO form is to find a bit-to-symbol function that represents the transmission symbols using the bit string. As mentioned before, this function directly influences the final detection accuracy. In this work, we use the bit-to-symbol mapper proposed in 5G NR standard \cite{3gpp2018}. For phase shift keying (PSK) modulation, this mapper is equivalent to ML-to-QUBO reduction in \cite{kim2019leveraging} but is quite different in quadrature amplitude modulation (QAM).

For $|\Omega|$-QAM, the number of bits to represent each symbol is $m=\log_2|\Omega|$, then the total number of bits for $N_r\times N_t$ MIMO system is $N_r\times m$. The ML-to-QUBO transformation is 
\begin{align}
&x_i(\{q_{mi-a}\}_{a=1}^m)=x_i^R+jx_i^I\notag\\
=&\underbrace{\sum_{a=1}^{m/2}2^{\frac{m}{2}+1-a}q_{mi-a}-(\frac{m}{2}-1)}_{\text{real part}}\notag\\
&+j\underbrace{\sum_{a=1}^{m/2}2^{\frac{m}{2}+1-a}q_{mi-\frac{m}{2}-a}-(\frac{m}{2}-1)}_{\text{image part}},
    \label{equ:qubo}
\end{align}
while the Gray-coded bit-to-symbol function is 
\begin{align}
    &x_i(\{q_{mi-a}\}_{a=1}^m)=x_i^R+jx_i^I\notag\\
    =&\underbrace{-\sum_{a=1}^{m/2}2^{\frac{m}{2}-a}\prod_{b=1}^aq_{mi-b}}_{\text{real part}}\notag\\
    &-j\underbrace{\sum_{a=1}^{m/2}2^{\frac{m}{2}-a}\prod_{b=1}^aq_{mi-\frac{m}{2}-b}}_{\text{image part}}, \quad i=1,\dots, N_r.\label{equ:qam}
\end{align}
It's worth noting that distinct bit-to-symbol functions can yield identical constellation while producing divergent loss landscapes, consequently leading to varying levels of optimization complexity.
ML-to-QUBO reduction method has shown promising performance in large MIMO systems in low-order modulation by using quantum annealing or quantum-inspired optimization methods~\cite{kim2021heuristic, kim2020towards,10.1145/3447993.3448619,singh2022ising} and thus, we will only discuss and compare the different bit-to-symbol functions for QAM in this paper.

Substituting bit-to-symbol function (Eq. \eqref{equ:qam}) into the objective function of ML (Eq. \eqref{equ:ml}), we can get the real-valued objective function in HUBO form (Eq. \ref{equ:hubo_mld}).
\begin{widetext}
\begin{align}
    &\min_{\bm{q}} \ 2\sum_{i=1}^{N_t}\sum_{j=1}^{N_r}\left[(H_{ij}^Ry_j^R+H_{ij}^Iy_j^I)x_i^R+(H_{ij}^Ry_j^I+H_{ij}^Iy_j^R)x_i^I\right]\notag\\
    &+2\sum_{i=1}^{N_r}\sum_{j=1}^{N_t}\sum_{k=j+1}^{N_t}\left[(H_{ij}^RH_{ik}^R+H_{ij}^IH_{ik}^I)(x_j^Rx_k^R+x_j^Ix_k^I)+(H_{ij}^RH_{ik}^I-H_{ij}^IH_{ik}^R)(x_j^Ix_k^R-x_j^Rx_k^I)\right]\notag\\
    &+\sum_{i=1}^{N_t}\sum_{j=1}^{N_r}({H_{ij}^R}^2+{H_{ij}^I}^2)({x_i^R}^2+{x_i^I}^2)\notag\\
    &=\min_{\bm{q}} \ -2\sum_{i=1}^{N_t}\sum_{j=1}^{N_r}\left[(H_{ij}^Ry_j^R+H_{ij}^Iy_j^I)\left(\sum_{a=1}^{m/2}2^{\frac{m}{2}-a}\prod_{b=1}^aq_{mi-b}\right)+(H_{ij}^Ry_j^I+H_{ij}^Iy_j^R)\left(\sum_{a=1}^{m/2}2^{\frac{m}{2}-a}\prod_{b=1}^aq_{mi-\frac{m}{2}-b}\right)\right]\notag\\
    &+2\sum_{i=1}^{N_r}\sum_{j=1}^{N_t}\sum_{k=j+1}^{N_t}\left[(H_{ij}^RH_{ik}^R+H_{ij}^IH_{ik}^I)\left(\sum_{a=1}^{m/2}\sum_{b=1}^{m/2}2^{m-a-b}\right.(\prod_{c=1}^a\prod_{d=1}^{b}q_{mj-c}q_{mk-d}+\prod_{c=1}^a\prod_{d=1}^{b}q_{mj-\frac{m}{2}-c}q_{mk-\frac{m}{2}-d})\right)\notag\\
     &+(H_{ij}^RH_{ik}^I-H_{ij}^IH_{ik}^R)\left.\left(\sum_{a=1}^{m/2}\sum_{b=1}^{m/2}2^{m-a-b}(\prod_{c=1}^a\prod_{d=1}^{b}q_{mj-\frac{m}{2}-c}q_{mk-d}-\prod_{c=1}^a\prod_{d=1}^{b}q_{mj-c}q_{mk-\frac{m}{2}-d})\right)\right]\notag\\
    &+\sum_{i=1}^{N_t}\sum_{j=1}^{N_r}({H_{ij}^R}^2+{H_{ij}^I}^2)\left[\sum_{k=1}^{m/2-1}\sum_{l=k+1}^{m/2}2^{m-k-l}\left(\prod_{a=k}^{l-1}q_{mi-a}+\prod_{a=k}^{l-1}q_{mi-\frac{m}{2}-a}\right)\right].
    \label{equ:hubo_mld}
\end{align}
\end{widetext}

The generated HUBO problem consists of three parts and it should be noted that the last part has been simplified since $q_i^2=1 (\forall i)$ which is important for optimization.

\subsection{Solver}
When using quantum annealer or Ising machines to solve HUBO problems, the problems always are transformed into QUBO by introducing auxiliary spin variables. However, p-bits is able to solve it directly~\cite{borders2019integer}. Next, we will briefly introduce the mechanism of p-bits and the simulated algorithm of it.


\subsubsection{P-bits}

P-bit is a new unconventional computation scheme that utilizes probabilistic bits which is intermediate between classical bit and qubit. And it has been demonstrated that p-bits can efficiently solve combinatorial optimization problems by accelerating randomized algorithms~\cite{kaiser2021probabilistic}. 

Every individual bit in p-bits is built with a spintronic device named magnetic tunnel junction (MTJ). According to tunnelling magnetoresistance effect, MTJs exhibit two different resistance states of high and low due to the relative direction of the two ferromagnetic layers, which can be used to represent binary variables. And MTJs are unstable that they are always fluctuating between the two states due to their intrinsic randomness. Arrhenius'law tells us that the expected time between two switches is about $\tau = \tau_0 \exp (\Delta E/{k_B T}),\tau_0\approx 1\text{ns}$. 

\subsubsection{Simulated p-bits algorithm}
In order to understand and predict the performance of p-bits on MIMO detection problems, we implemented a simulator of high-order p-bits~\cite{nikhar2023all}, namely Simulated P-bit algorithm, based on the mechanism of p-bits.

By analysing of stochastic processes of p-bits, we can calculate the switch probabilities of each bit. The only parameter that matters is $\Delta E$ so that the switch probabilities can be finely controlled by its value. In p-circuits, $\Delta E$ is tuned by the current signal automatically. In other words, a p-bit can be described as a binary stochastic neuron in mathematics,
\begin{equation}
    m_i = \text{sgn}[\tanh (\beta I_i) - r],
    \label{equ: m}
\end{equation}
where $r\sim\mathcal{U}(-1, 1)$ represents a random number. $I_i$ is the input signal, a continuous value, and $m_i$ is the output signal with binary value. $\beta$ is the inverse temperature, which increases during the annealing. The circuit that calculates the input signal encodes a special problem. For a HUBO problem, $I_i$ should be
\begin{equation}
    I_i = -\frac{\partial H(\bm{m})}{\partial m_i}.
    \label{equ:I}
\end{equation}
where $H(\bm{m})$ is a hamiltonian of HUBO problem. When we run p-bits to solve a combinatorial optimization problem, actually it is going through an annealing process which helps algorithm to escape from local minima and converge to the solution with lower energy. The most difference between P-bit and traditional simulated annealing is that the former can update spins in parallel while the latter serially update the spins~\cite{onizawa2024enhanced}. The Simulated p-bits algorithm is described in Algorithm \ref{al:p-bit}.


\begin{algorithm}
\caption{Simulated P-bit}
\label{al:p-bit}
\KwIn{Initial replicas with random states}
\KwOut{Final states of the P-bits}
$\beta \gets \beta_0$\;
\For{each trial}{
    \For{each p-bit}{
        Compute input signal $I_i = -\frac{\partial H(\bm{m})}{\partial m_i}$\;
        Compute the output state $m_i = \text{sgn}[\tanh (\beta I_i) - r]$\;
    }
    $\beta \gets \beta + \Delta_\beta$\;
}
\end{algorithm}

The value of input signal $I_i$ is computed by traversing the neighbours of spin $i$ and this operation accounts for the majority of the computation time for each spin update. In practice, the average probability of a spin flip is typically around 15\% for most annealing schedules. Additionally, there are numerous shared terms in the input signals across different spins, which could be reused to reduce redundant calculations. Therefore, it is more efficient to update the corresponding terms only when the spin is flipped. 

\subsection{The pipeline of HOPbit}

\begin{figure*}
	\centering
\includegraphics[width=0.8\linewidth]{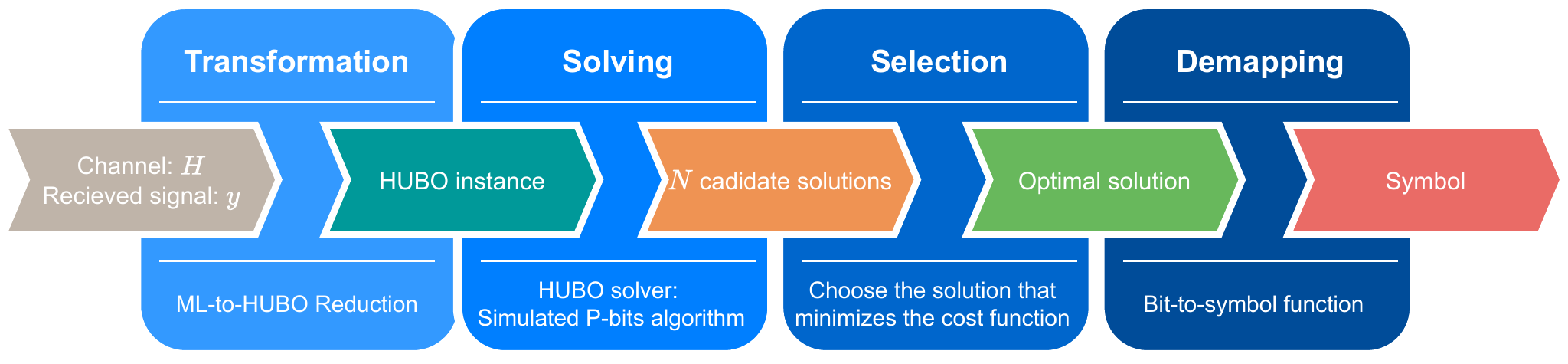}
	\caption{The flowchart of HOPbit.}
	\label{fig: HuMax}
\end{figure*}
The pipeline of HOPbit contains three parts, including transformation, solving, selection, and demapping (Fig.~\ref{fig: HuMax}). Specifically, it is listed as follows:
\begin{itemize}
    \item Reduce MIMO detection problem into the HUBO form;
    \item Perform $N$ runs of simulated p-bits algorithm to get a set of candidate solutions;
    \item Select the optimal solution from the candidate solutions based on Eq.~(\ref{equ:H}).
    \item Directly transform bits into symbols based on Eq.~(\ref{equ:qam}).
\end{itemize}

\section{Experiments}
To investigate the effectiveness of HOPbit, some evaluations and comparisons with other state-of-the-art detectors are conducted on large MIMO systems. In this section, we first present the experimental dataset, and then introduce evaluation metrics, and finally present the results.

\subsection{Experimental setup}
We simulate wireless MIMO channels between users and receiver antennas through independent and identically distributed (i.i.d) Gaussian channels as well as additive white Gaussian noise (AWGN) for SNR ranging from 15dB to 55dB. Each SNR setting contains 100 different instances in order to assess the average performance of the detectors. The MIMO sizes are 4$\times 4$ for high-order modulations including 16-QAM, 64-QAM and 256-QAM. In order to verify the ability of HOPbit to eliminate the inconsistency error, we compare HOPbit with the detector based on ML-to-QUBO transformation and use the same solver as HOPbit, which is denoted as QOPbit here. In addition, some other state-of-the-art detectors in large and massive MIMO systems, including MMSE, LArge MIMO Approximate message passing (LAMA)~\cite{jeon2015optimality, bitra2022large} and the optimal Sphere Decoder(SD)-based maximum likelihood MIMO detector~\cite{burg2005vlsi}, are included in the comparison. LAMA iteratively estimates the probability distribution of signals based on approximate message passing. It has low complexity and is readily implementable in practice, but it works well only in the specific scenario~\cite{he2023gnn}. SD essentially searches for the most likely transmitted symbol by considering a hypersphere in the signal space and iteratively refining the search space based on pruned tree search~\cite{hung2006sphere}. However, it still suffers from exponentially increasing in computational demand.

To further show the advantage of HOPbit, we implement the same experiment setting as ParaMax and compare the HOPbit with ZF, ParaMax and 2R-ParaMax in 12-user 16-QAM. That is, we set the number of iterations to 50.

To evaluate the performance of HOPbit in more realistic channel models, we compare HOPbit with QOPbit as well as some traditional detectors (e.g. ZF, LMMSE) in flat fading MIMO channels. We simulate $N_t=16$ MIMO systems with QAM-16, QAM-64 and QAM-256 for various values of $N_r$ ($N_r=16, 64$) at ranging from 5 to 30 dB SNR by Sionna~\cite{hoydis2022sionna}, a Python library. For each setting,  we randomly generate 100,000 instances in order to achieve BER up to near $10^{-6}$.

\begin{figure*}
	\centering
	\subfloat[16-QAM 4$\times$4]
	{\includegraphics[width=0.45\linewidth]{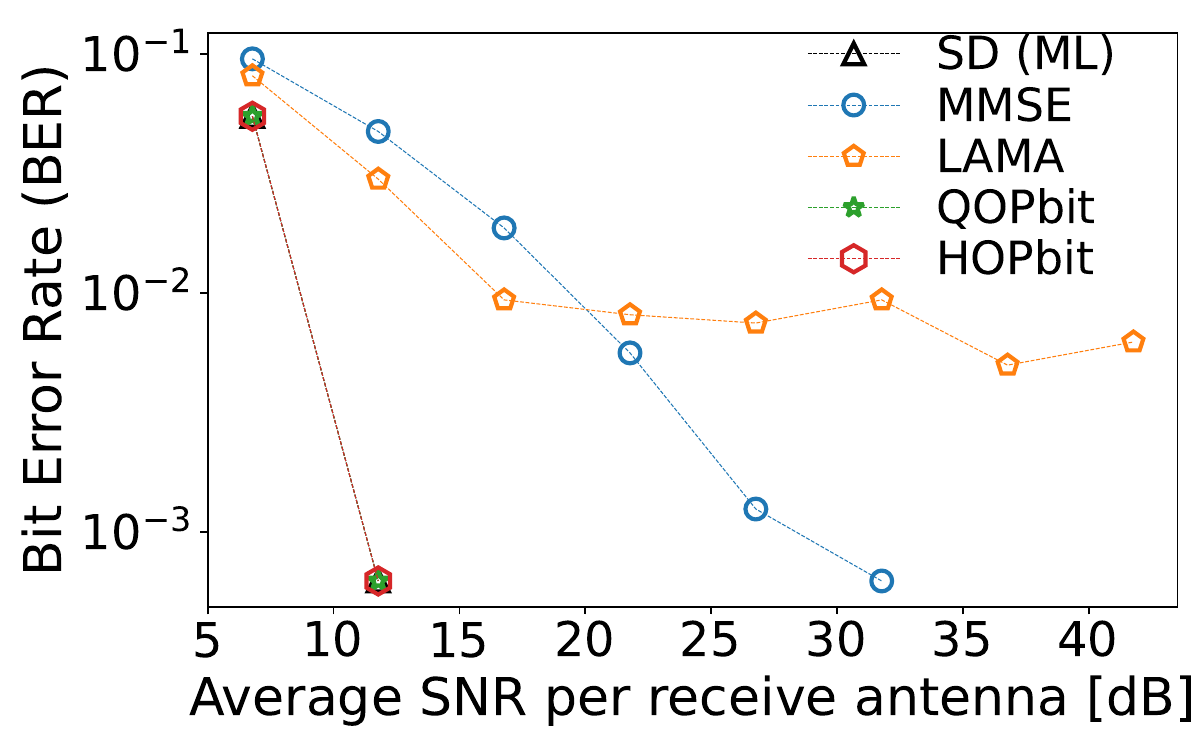}}
    \subfloat[64-QAM 4$\times$4]
	{\includegraphics[width=0.45\linewidth]{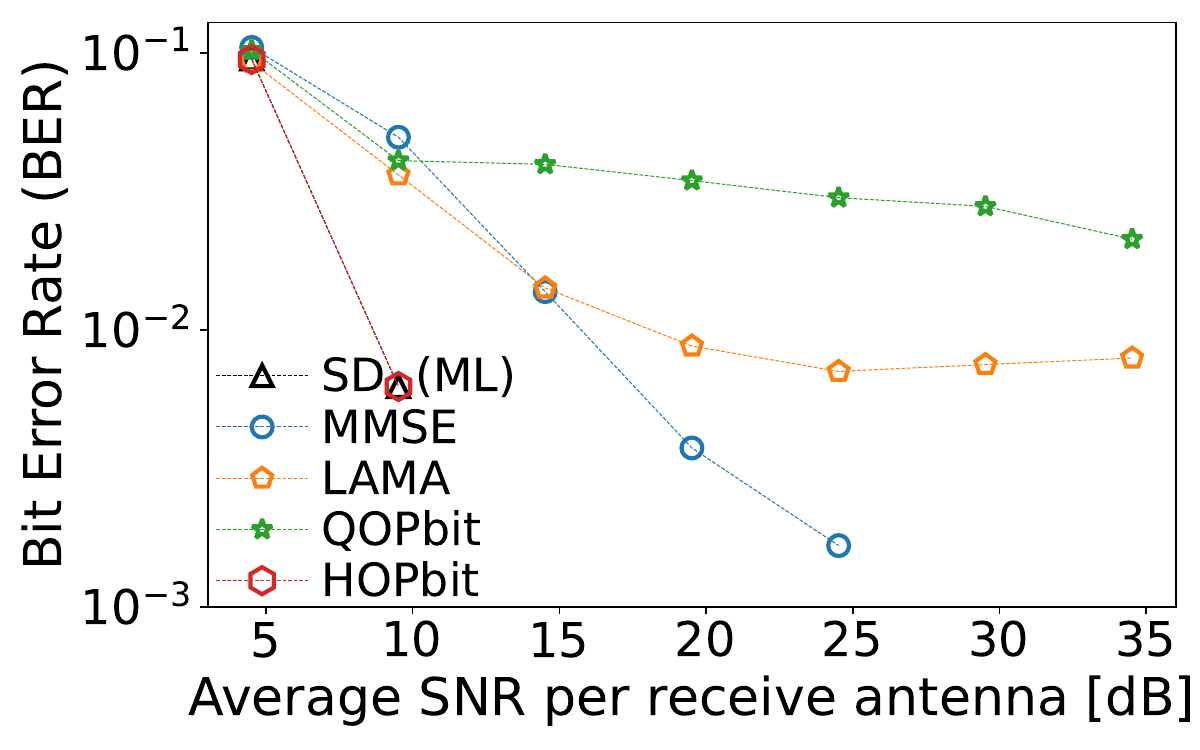}}
    
    \subfloat[256-QAM 4$\times$4]
	{\includegraphics[width=0.45\linewidth]{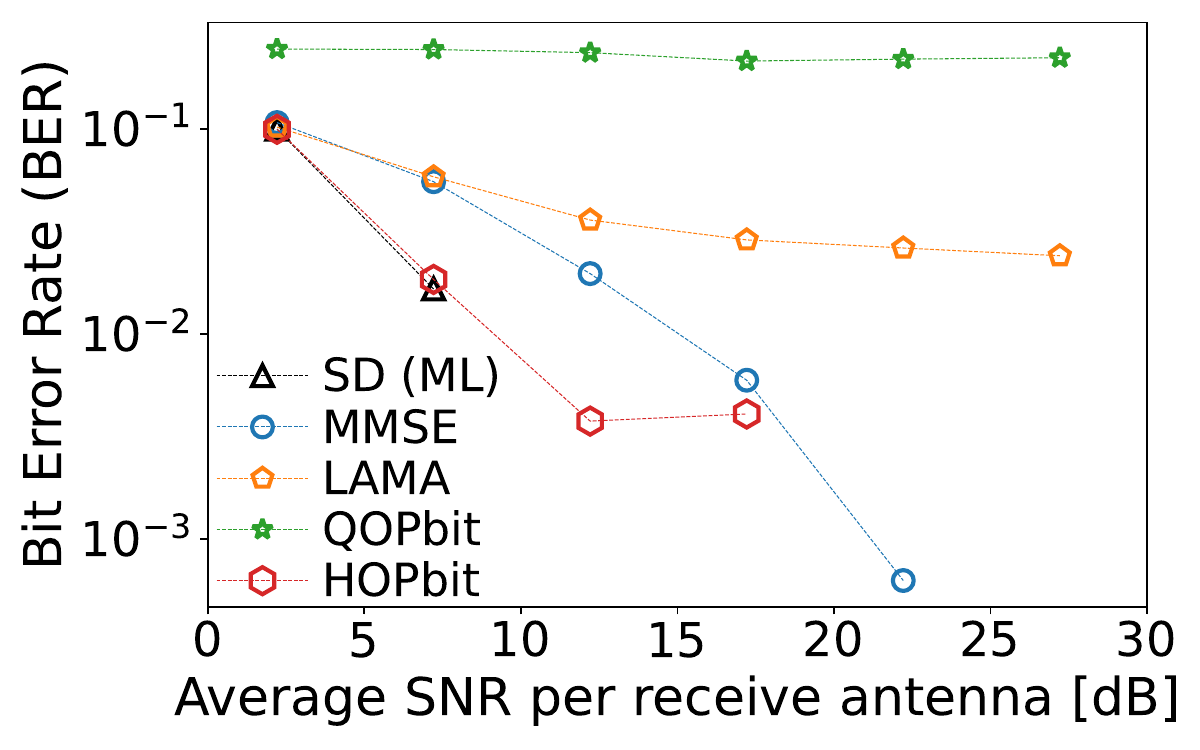}}
	\caption{Comparisons of BER of different detectors on different large MIMO systems with high-order modulations on Gaussian channel. The missing data points mean no error for detectors within the experimental instances.}
	\label{fig: result}
\end{figure*}

\subsection{Results}
The detection accuracy is measured by BER, the ratio of the bits that have been detected by mistake relative to the total number of bits transmitted in a transmission. The number of iterations and anneals of HOPbit and QOPbit are the same, 30 and 100 respectively, in order to compare detectors fairly. QOPbit exhibits inferior performance compared to HOPbit (Fig.~\ref{fig: result}). The performance of QOPbit deteriorates significantly with increasing modulation order, as the inconsistency error accumulates. In contrast, HOPbit leverages ML-to-HUBO transformation to mitigate the inconsistency error, enabling rapid convergence to an approximate optimal solution. Consequently, HOPbit also outperforms MMSE and LAMA, and exhibits a similar detection accuracy as SD. As a result, it further extends QIA-based detectors' feasible MIMO regimes. The relatively diminished performance of HOPbit at high SNR arises from its fixed parameter setting for all instances with the same MIMO size (Fig~\ref{fig: result}). It is reasonable to tailor the annealing schedule according to the noise level to enhance performance. 

\begin{figure}
	\centering
\includegraphics[width=0.8\linewidth]{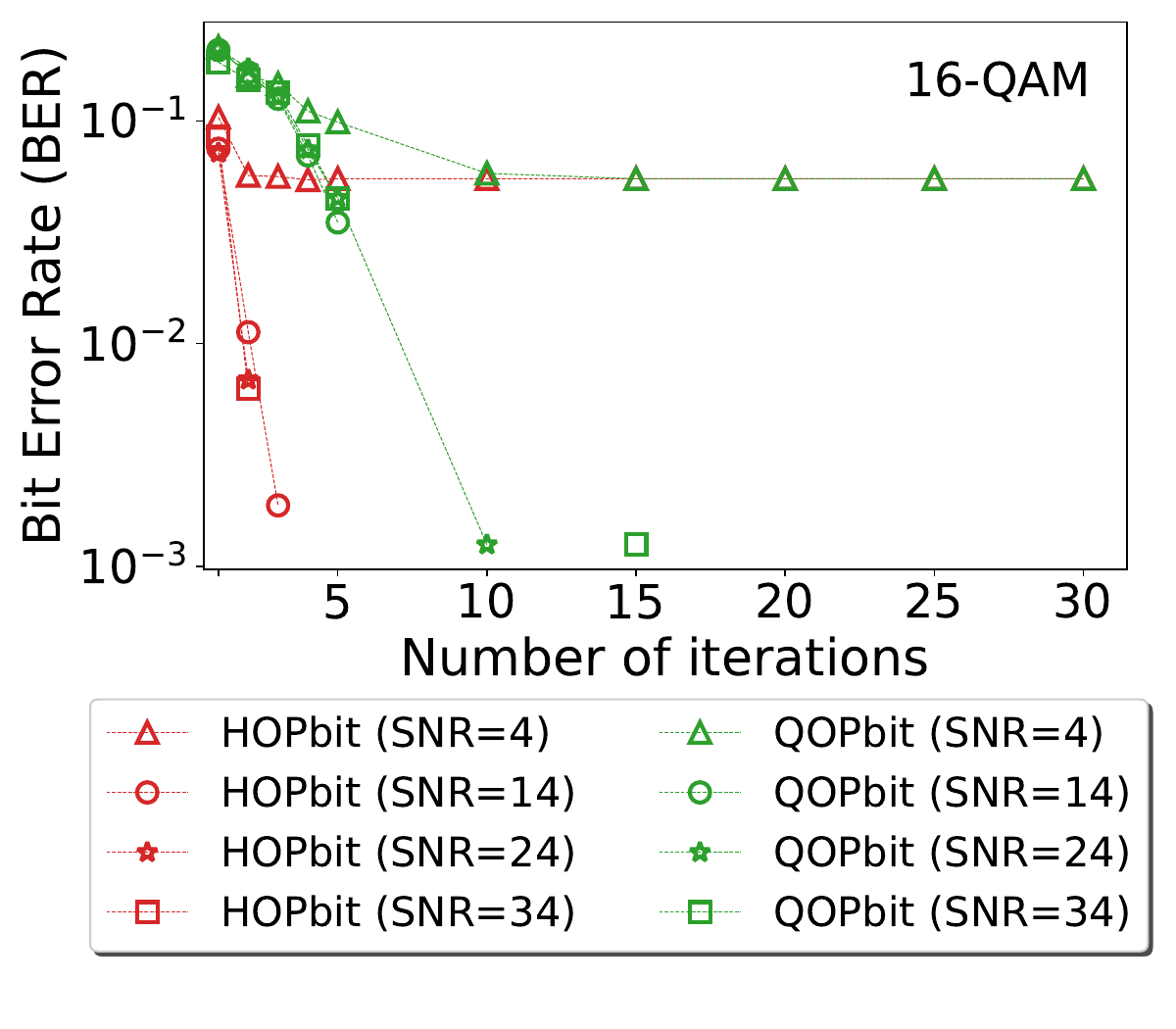}
	\caption{BER as a function of the number of iterations for HOPbit and QOPbit detectors in four-user 16-QAM MIMO systems with varying high-order modulations. The missing data points mean no error for detectors within the experimental instances.}
	\label{fig: iter}
\end{figure}

Although HOPbit and QOPbit reach the same BER in four-user 16-QAM MIMO systems, we evaluate BER as a function of iterations for them and find that HOPbit converges extremely fast, with only three iterations approximately across varying SNR settings while QOPbit converges after 15-20 iterations (Fig.~\ref{fig: iter}). It further illustrates that ML-to-QUBO transformation can accelerate the convergence of detectors.

\begin{figure}
	\centering
\includegraphics[width=0.8\linewidth]{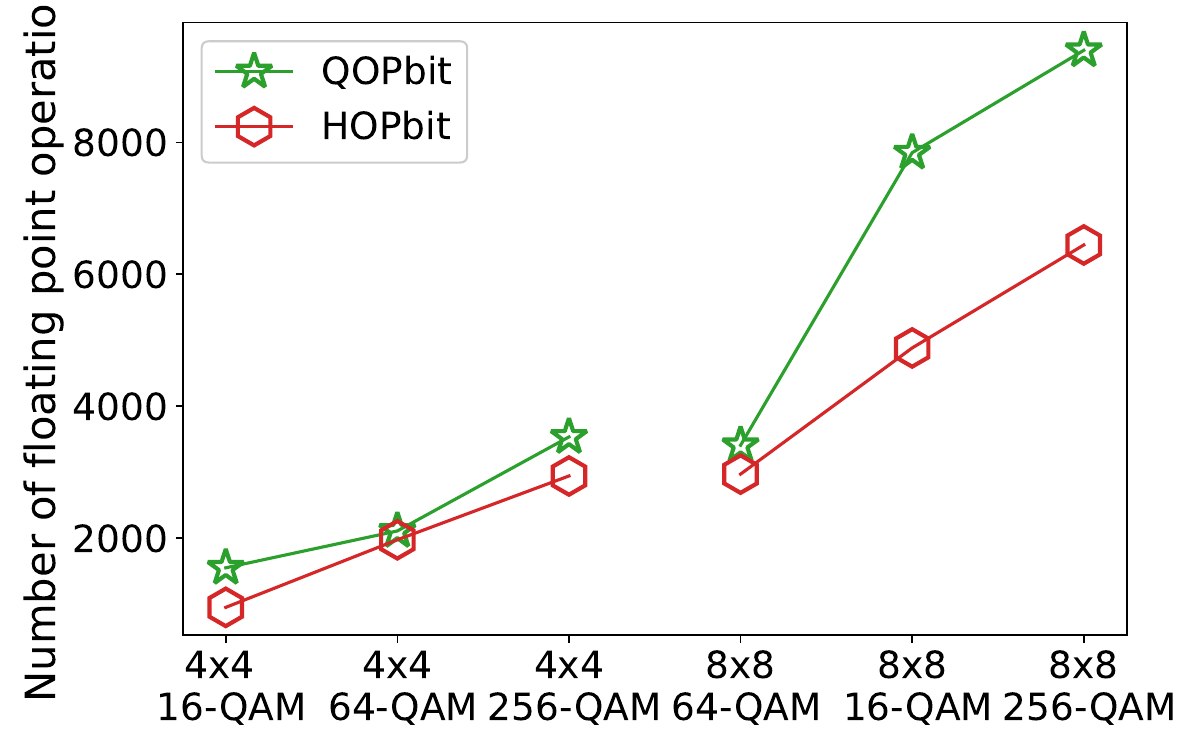}
	\caption{Comparison of computational complexity between QOPbit and HOPbit with precomputation in varying MIMO systems.}
	\label{fig: flop}
\end{figure}

Then, we analyze the computational complexity of MMSE, QOPbit and HOPbit with and without pre-computation (Table ~\ref{tab:complexity}). With pre-computation, QOPbit and HOPbit can only compute the terms related to the flipped bits while keeping other terms unchanged, which can further reduce the computational complexity. For MMSE, the computational complexity for optimizing is quite low after pre-processing the channel matrix. For QOPbit and HOPbit, the computational overhead of pre-processing is low, but the optimization is time-consuming. To sum up, the time complexity of MMSE and QOPbit/HOPbit is $O(N_r^2N_t)$ and $O(N_rN_t^2)$, respectively, which means that QOPbit and HOPbit are much more efficient than MMSE. The computation complexity of QOPbit and HOPbit is quite close in a single iteration. Taking the faster convergence of HOPbit into account, the whole processing time of HOPbit is much lower than QOPbit, especially for a larger system with higher modulations (Fig.~\ref{fig: flop}). On the other hand, parallel computing can benefit QOPbit and HOPbit, that is, it can improve the performance and reduce the execution time while MMSE is not suitable for parallel computing. Equipped with enough processing elements (the number of processing elements is no more than the number of anneals in this work), the computational time of HOPbit with precomputation is enough for LTE requirements.
\begin{table*}[]
\centering 
\caption{Computational complexity analysis for MMSE, QOPbit and HOPbit with and without pre-computation. $I$ is the number of iterations and $\tilde{I}(\ll 2IN_tm)$ denotes the number of updated bits during the iterations.}
\begin{tabular}{@{}lllll@{}}
\toprule
\multirow{2}{*}{Methods}  & \multicolumn{2}{c}{Pre-processing (update $H$)} & \multicolumn{2}{c}{Optimization} \\ \cmidrule(l){2-5} 
                          & multiplication      & addition      & multiplication   & addition   \\ \cmidrule(r){1-5}
\multicolumn{1}{l}{\multirow{4}{*}{MMSE}} &  \multirow{4}{*}{\begin{tabular}{l}$24N_r^2N_t+\frac{8}{3}N_t^3$\\$+3N_t^2-\frac{13}{6}N_t$\end{tabular}}                    &   \multirow{4}{*}{\begin{tabular}{l}$24N_r^2N_t+\frac{8}{3}N_t^3$\\$+2N_t^2-4N_r^2$\\$-8N_tN_r+2N_r$\\$-\frac{5}{6}N_t$ \end{tabular}}           &    $4N_rN_t+2N_t$             & $2N_t(2N_r+2^\frac{m}{2}-2)$            \\
                          &                     &              &                  &            \\
                          &                     &              &                  &            \\
                          &                     &              &                  &            \\
                        &                     &              &                  &            \\
\multicolumn{1}{l}{\multirow{3}{*}{QOPbit}}                          &  $8N_rN_t^2$                   &      $8N_rN_t^2-4N_t^2$        & \multirow{2}{*}{\begin{tabular}{l}$2ImN_t^2+8N_rN_t+$\\$(\frac{Im^2}{2}+6Im+m-2)N_t$\end{tabular}}                 &       \multirow{2}{*}{\begin{tabular}{l}$2ImN_t^2+8N_rN_t+$\\$(\frac{Im^2}{2}+5Im+m-4)N_t$\end{tabular}}     \\
                          &                     &              &                  &            \\
                          &                     &              &                  &            \\
                          &                     &              &                  &            \\
\multicolumn{1}{l}{\multirow{3}{*}{HOPbit}}                          &   $8N_rN_t^2$                   &       $8N_rN_t^2-4N_t^2$        &   \multirow{2}{*}{\begin{tabular}{l}$2ImN_t^2+8N_rN_t+$\\$(7Im+m-2)N_t$\end{tabular}}               &     \multirow{2}{*}{\begin{tabular}{l}$2ImN_t^2+8N_rN_t+$\\$(5Im+m-4)N_t$\end{tabular}}       \\ 
                          &                     &              &                  &            \\
                          &                     &              &                  &            \\
                          &                     &              &                  &            \\
\multicolumn{1}{c}{\multirow{3}{*}{\begin{tabular}{l}QOPbit\\ precomputation\end{tabular}}}                        &   $8N_rN_t^2$                   &       $8N_rN_t^2-4N_t^2$        &   \multirow{2}{*}{\begin{tabular}{l}$4N_rN_t+2mN_t^2+$\\$(2Im+6m)N_t+2\tilde{I}$\end{tabular}}               &     \multirow{2}{*}{\begin{tabular}{l}$4N_rN_t+2mN_t^2+$\\$(2Im+2\tilde{I}+m-2)N_t$\end{tabular}}       \\ 
                          &                     &              &                  &            \\
                          &                     &              &                  &            \\
                          &                     &              &                  &            \\
\multicolumn{1}{c}{\multirow{4}{*}{\begin{tabular}{l}HOPbit\\ precomputation\end{tabular}}}                        &   $8N_rN_t^2$                   &       $8N_rN_t^2-4N_t^2$        &   \multirow{3}{*}{\begin{tabular}{l}$4N_rN_t+2mN_t^2+$\\$(2Im+6m)N_t+\frac{m}{2}\tilde{I}$\end{tabular}}               &     \multirow{3}{*}{\begin{tabular}{l}$4N_rN_t+2mN_t^2+$\\$(2Im+2\tilde{I}+m-2)N_t$\\$+(\frac{m}{2}-2)\tilde{I}$\end{tabular}}       \\ 
                          &                     &              &                  &            \\
                          &                     &              &                  &            \\
                          &                     &              &                  &            \\
                          &                     &              &                  &            \\ 
\bottomrule
\end{tabular}
\label{tab:complexity}
\end{table*}

\begin{figure}
	\centering
\includegraphics[width=1.\linewidth]{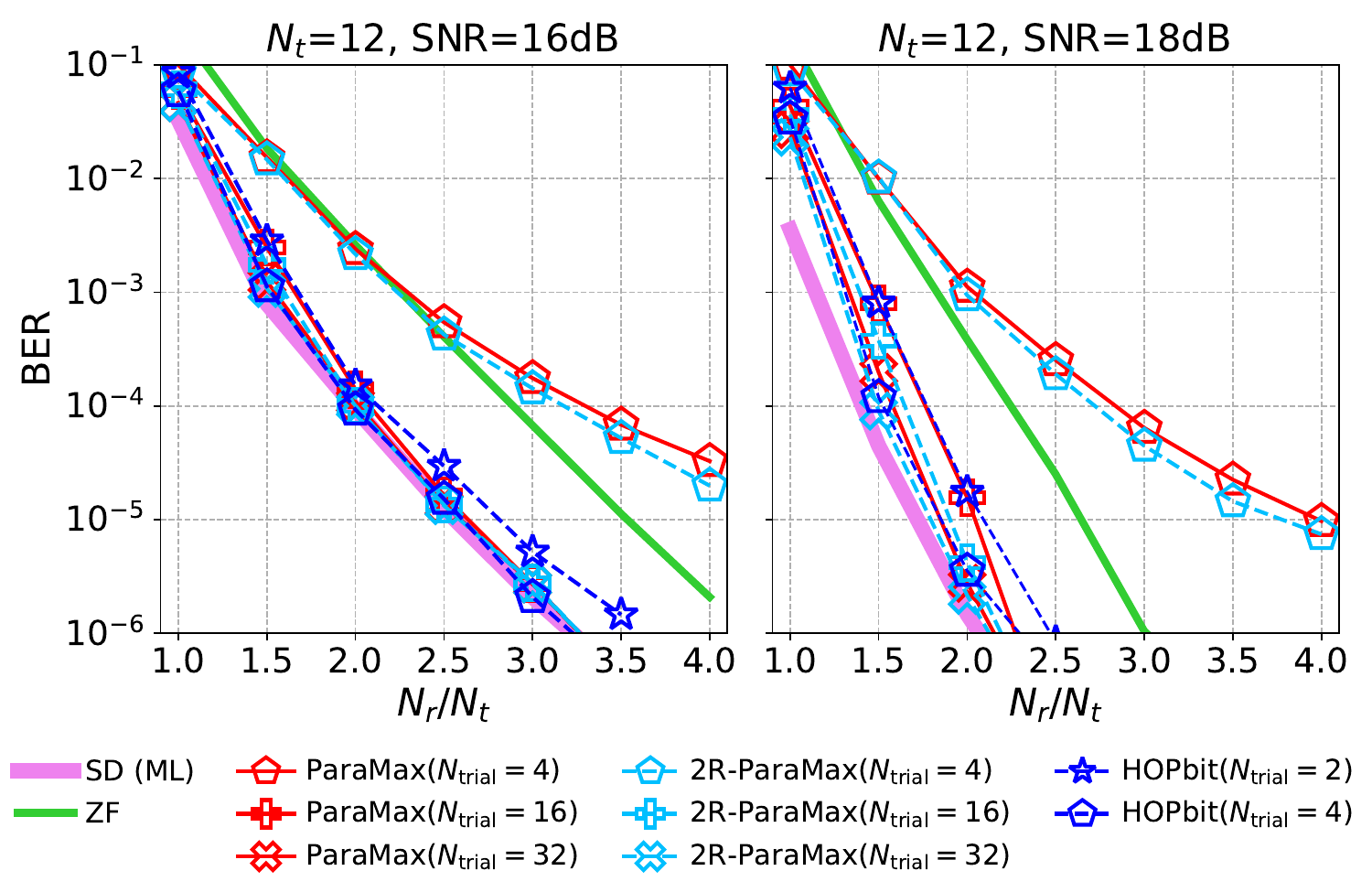}
	\caption{Comparisons of BER for HOPbit and ZF in 12-user MIMO regimes and/or SNRs with 16-QAM. It is worth noting that the results of SD, ParaMax and 2R-ParaMax are obtained from literatures~\cite{10.1145/3447993.3448619}.}
	\label{fig: NrNt_ber}
\end{figure}

\begin{figure}
	\centering
\includegraphics[width=1.\linewidth]{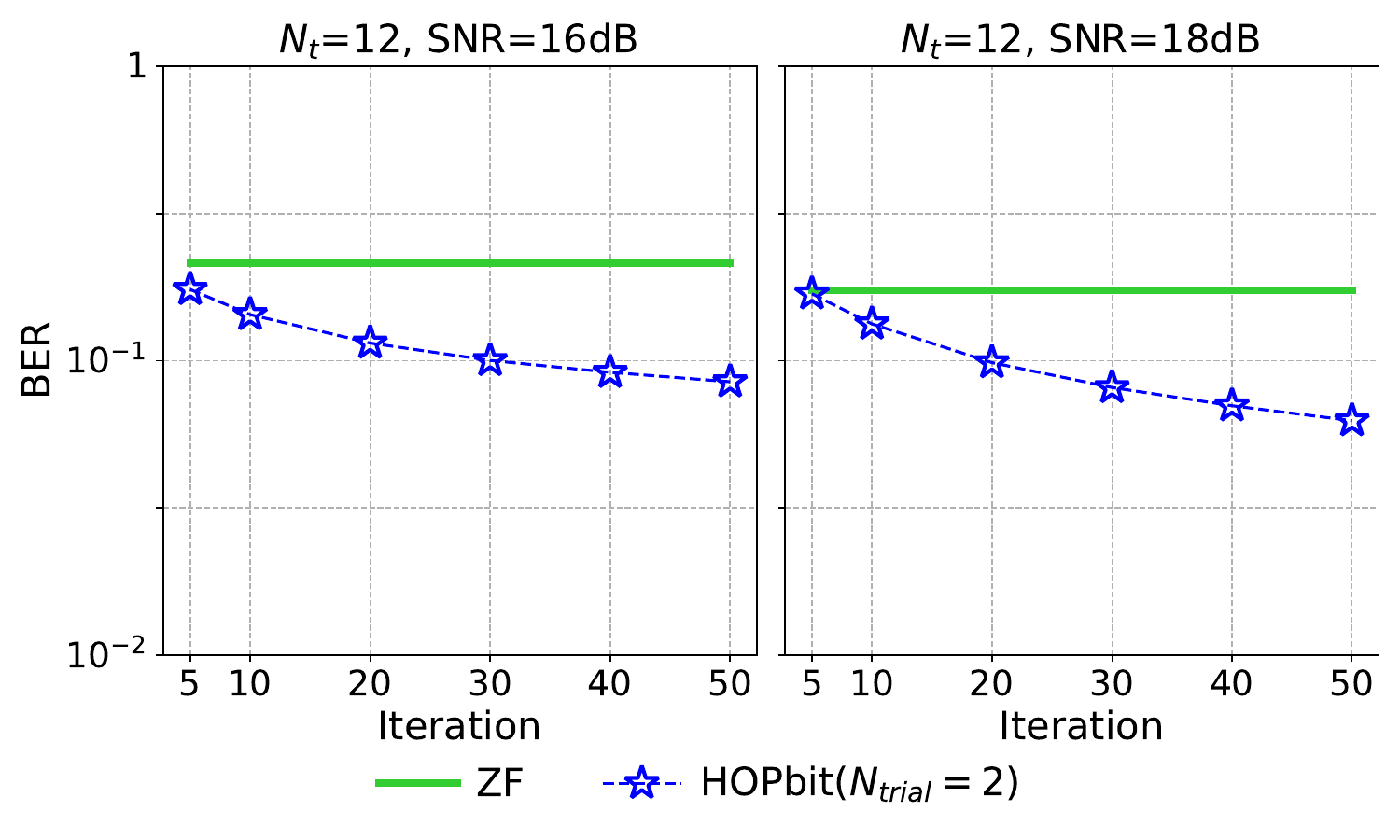}
	\caption{BER as a function of $N_{\text{trial}}$ varying SNRs in 12$\times$12 large MIMO.}
	\label{fig: iter_ber}
\end{figure}

Besides, we evaluate the BER of ZF and HOPbit as well as the results from literature~\cite{10.1145/3447993.3448619} in various 12-user 16-QAM MIMO systems. The BER curve of ZF in Fig.~\ref{fig: NrNt_ber} is almost the same as shown in the previous paper~\cite{10.1145/3447993.3448619}, which further illustrates the consistency of the experimental setup. Both ParaMax and 2R-ParaMax underperform by several orders of magnitude BER in terms of Massive MIMO compared to ZF with 4 trials (each processing element contains 2 trials, so it adds up to 4 trials in 2 processing elements). Only when the number of trials increases to 16, the BER performance of Paramax becomes better than ZF and close to SD. For HOPbit, it can be observed that it outperforms ZF in all the 12-user MIMO regimes with 16-QAM even only with 2 trials. Equipped with 4 trials, HOPbit obtains the lower BER, which extremely approximates the performance of SD. Besides, thanks to the fast convergence of HOPbit, we only need 5-10 iterations with two trials to outperform ZF (Fig.~\ref{fig: iter_ber}). 

\begin{widetext}
\begin{figure*}
	\centering
	\subfloat[16-QAM 16$\times$16]
	{\includegraphics[width=0.49\linewidth]{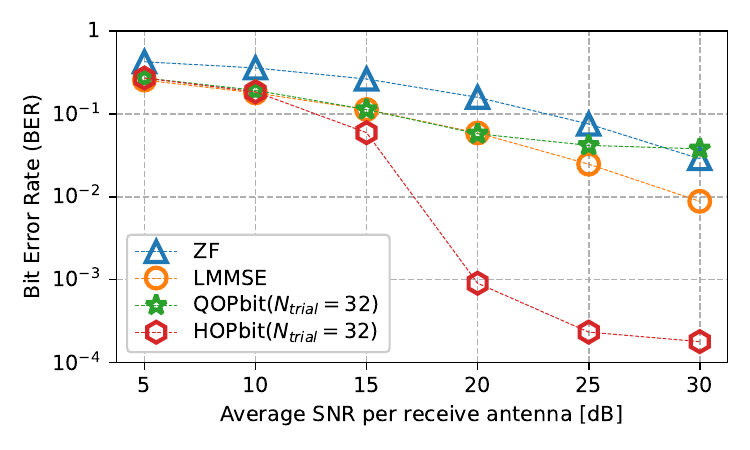}}
    \subfloat[16-QAM 16$\times$64]
    {\includegraphics[width=0.49\linewidth]{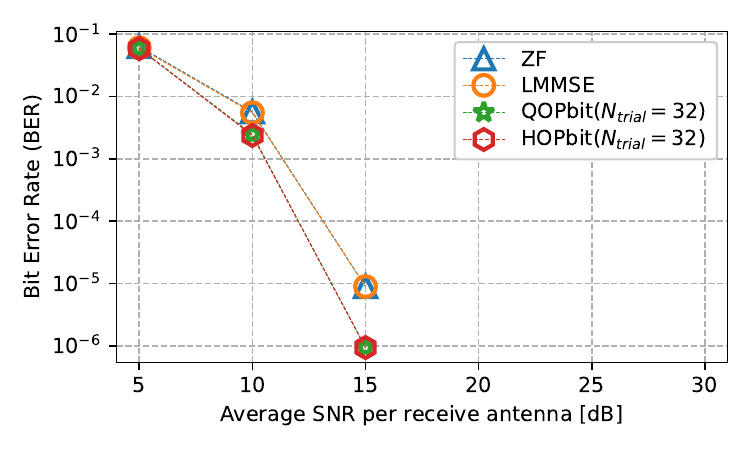}}
    
    \subfloat[64-QAM 16$\times$16]
	{\includegraphics[width=0.49\linewidth]{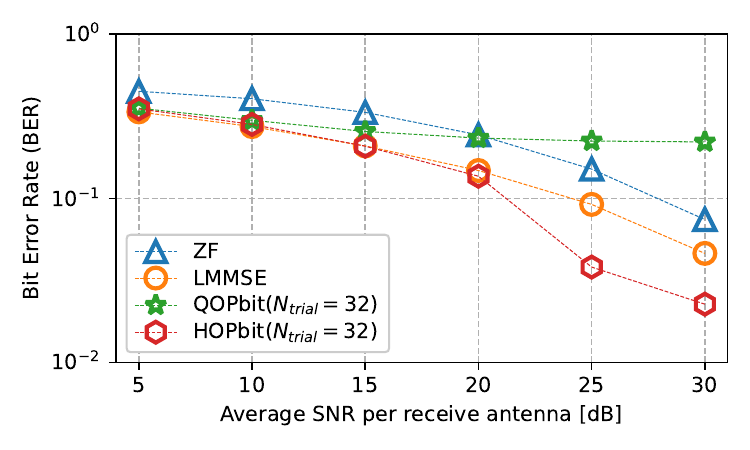}}
    \subfloat[64-QAM 16$\times$64]
    {\includegraphics[width=0.49\linewidth]{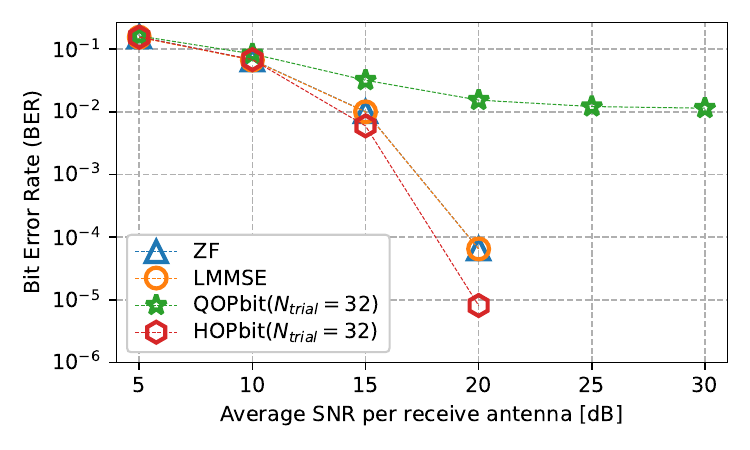}}
    
    \subfloat[256-QAM 16$\times$16]
	{\includegraphics[width=0.49\linewidth]{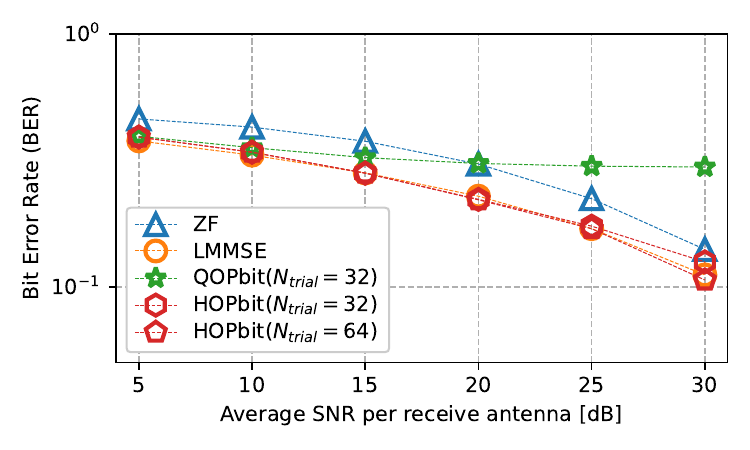}\label{fig: ff256qam16x16}}
    \subfloat[256-QAM 8$\times$8]
    {\includegraphics[width=0.49\linewidth]{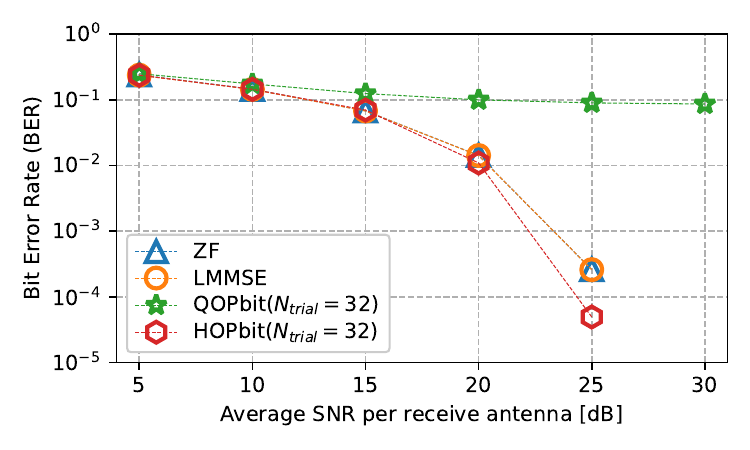}}
	\caption{Comparisons of BER of different detectors on different large MIMO systems with high-order modulations on flat fading channel. The missing data points mean no error for detectors within the experimental instances.}
	\label{fig: flatfading}
\end{figure*}
\end{widetext}

Finally, we evaluate the performance of various detectors on flat fading MIMO channels (shown in Fig.~\ref{fig: flatfading}). Due to the large size of MIMO systems, ML detectors in Sionna are unable to solve it. Therefore, the comparison is conducted among HOPbit, QOPbit, ZF, and LMMSE. The conclusion is the same as that in Gaussian channels. QOPbit only performs well as HOPbit in small MIMO systems with low order modulations. As the order of modulations or the size of MIMO systems increases, QOPbit suffers from error floor and performance deteriorates, even worse than ZF and LMMSE. Although equipped with more trials, it cannot overcome error floor. However, HOPbit solves error floor and always achieves the best performance among all the detectors we involved. In Fig.~\ref{fig: ff256qam16x16}, we increase the number of trial to improve the performance of HOPbit but it will not increase the computing time since it can be processed in parallel.


\section{Conclusion}
Previous research endeavors have revealed the challenges that quantum and physics-inspired methods encounter when dealing with high-order modulations, even when subjected to increased iterations and trials. In response, this study introduces the HOPbit detector, a novel approach designed to address these issues. HOPbit first leverages the ML-to-HUBO transformation and then employs the simulated p-bits algorithm to effectively solve the large and massive MIMO detection problem. 

Remarkably, the HOPbit detector demonstrates swift convergence and achieves near-maximum-likelihood performance, even in the presence of high-order modulation schemes within the context of large MIMO systems. Experimental results confirm that HOPbit not only outperforms QUBO solvers, including QOPbit, ParaMax and 2R-ParaMax configured with identical settings, but also surpasses traditional detectors such as ZF, MMSE, and LAMA.

However, there are still several challenges or improvements that can be discussed here.
\begin{itemize}
    \item \textbf{Parameters setting: }The performance of HOPbit is sensitive to the parameters, including annealing schedule-related and the number of iterations, especially for the higher-order modulations and large MIMO systems. Therefore, it is necessary to develop an automatic parameter tuning method for annealing-based detectors. In addition, the parameters should be set according to the noise level, like MMSE and LAMA.
    \item \textbf{Connection between different anneals: }The anneals in this work is independent, which means that the final results rely severely on the initialization. Simple repeated sampling may result in the waste of computing resources. We can solve it by learning from parallel tempering~\cite{swendsen1986replica}, reverse annealing~\cite{venturelli2019reverse} and tabu search~\cite{beasley1998heuristic}, etc.
    \item \textbf{Adaptability to Emerging Hardware:} Although HOPbit does not require any specialized hardware, it is adaptable with the emerging p-bits solvers~\cite{nikhar2023all}. In addition, it is also compatible with other specialized hardware, such as CMOS-based high-order Ising machine~\cite{su2023reconfigurable} and high-order oscillator Ising machine~\cite{bybee2023efficient}.
\end{itemize}

\nocite{*}


\begin{thebibliography}{55}%
\makeatletter
\providecommand \@ifxundefined [1]{%
 \@ifx{#1\undefined}
}%
\providecommand \@ifnum [1]{%
 \ifnum #1\expandafter \@firstoftwo
 \else \expandafter \@secondoftwo
 \fi
}%
\providecommand \@ifx [1]{%
 \ifx #1\expandafter \@firstoftwo
 \else \expandafter \@secondoftwo
 \fi
}%
\providecommand \natexlab [1]{#1}%
\providecommand \enquote  [1]{``#1''}%
\providecommand \bibnamefont  [1]{#1}%
\providecommand \bibfnamefont [1]{#1}%
\providecommand \citenamefont [1]{#1}%
\providecommand \href@noop [0]{\@secondoftwo}%
\providecommand \href [0]{\begingroup \@sanitize@url \@href}%
\providecommand \@href[1]{\@@startlink{#1}\@@href}%
\providecommand \@@href[1]{\endgroup#1\@@endlink}%
\providecommand \@sanitize@url [0]{\catcode `\\12\catcode `\$12\catcode
  `\&12\catcode `\#12\catcode `\^12\catcode `\_12\catcode `\%12\relax}%
\providecommand \@@startlink[1]{}%
\providecommand \@@endlink[0]{}%
\providecommand \url  [0]{\begingroup\@sanitize@url \@url }%
\providecommand \@url [1]{\endgroup\@href {#1}{\urlprefix }}%
\providecommand \urlprefix  [0]{URL }%
\providecommand \Eprint [0]{\href }%
\providecommand \doibase [0]{https://doi.org/}%
\providecommand \selectlanguage [0]{\@gobble}%
\providecommand \bibinfo  [0]{\@secondoftwo}%
\providecommand \bibfield  [0]{\@secondoftwo}%
\providecommand \translation [1]{[#1]}%
\providecommand \BibitemOpen [0]{}%
\providecommand \bibitemStop [0]{}%
\providecommand \bibitemNoStop [0]{.\EOS\space}%
\providecommand \EOS [0]{\spacefactor3000\relax}%
\providecommand \BibitemShut  [1]{\csname bibitem#1\endcsname}%
\let\auto@bib@innerbib\@empty
\bibitem [{\citenamefont {Albreem}\ \emph {et~al.}(2019)\citenamefont
  {Albreem}, \citenamefont {Juntti},\ and\ \citenamefont
  {Shahabuddin}}]{albreem2019massive}%
  \BibitemOpen
  \bibfield  {author} {\bibinfo {author} {\bibfnamefont {M.~A.}\ \bibnamefont
  {Albreem}}, \bibinfo {author} {\bibfnamefont {M.}~\bibnamefont {Juntti}},\
  and\ \bibinfo {author} {\bibfnamefont {S.}~\bibnamefont {Shahabuddin}},\
  }\bibfield  {title} {\bibinfo {title} {Massive mimo detection techniques: A
  survey},\ }\href@noop {} {\bibfield  {journal} {\bibinfo  {journal} {IEEE
  Communications Surveys \& Tutorials}\ }\textbf {\bibinfo {volume} {21}},\
  \bibinfo {pages} {3109} (\bibinfo {year} {2019})}\BibitemShut {NoStop}%
\bibitem [{\citenamefont {Trotobas}\ \emph {et~al.}(2020)\citenamefont
  {Trotobas}, \citenamefont {Nafkha},\ and\ \citenamefont
  {Lou{\"e}t}}]{trotobas2020review}%
  \BibitemOpen
  \bibfield  {author} {\bibinfo {author} {\bibfnamefont {B.}~\bibnamefont
  {Trotobas}}, \bibinfo {author} {\bibfnamefont {A.}~\bibnamefont {Nafkha}},\
  and\ \bibinfo {author} {\bibfnamefont {Y.}~\bibnamefont {Lou{\"e}t}},\
  }\bibfield  {title} {\bibinfo {title} {A review to massive mimo detection
  algorithms: Theory and implementation},\ }in\ \href@noop {} {\emph {\bibinfo
  {booktitle} {Advanced Radio Frequency Antennas for Modern Communication and
  Medical Systems}}}\ (\bibinfo  {publisher} {IntechOpen},\ \bibinfo {year}
  {2020})\BibitemShut {NoStop}%
\bibitem [{\citenamefont {Yang}\ and\ \citenamefont
  {Hanzo}(2015)}]{yang2015fifty}%
  \BibitemOpen
  \bibfield  {author} {\bibinfo {author} {\bibfnamefont {S.}~\bibnamefont
  {Yang}}\ and\ \bibinfo {author} {\bibfnamefont {L.}~\bibnamefont {Hanzo}},\
  }\bibfield  {title} {\bibinfo {title} {Fifty years of mimo detection: The
  road to large-scale mimos},\ }\href@noop {} {\bibfield  {journal} {\bibinfo
  {journal} {IEEE Communications Surveys \& Tutorials}\ }\textbf {\bibinfo
  {volume} {17}},\ \bibinfo {pages} {1941} (\bibinfo {year}
  {2015})}\BibitemShut {NoStop}%
\bibitem [{\citenamefont {Larsson}\ \emph {et~al.}(2014)\citenamefont
  {Larsson}, \citenamefont {Edfors}, \citenamefont {Tufvesson},\ and\
  \citenamefont {Marzetta}}]{larsson2014massive}%
  \BibitemOpen
  \bibfield  {author} {\bibinfo {author} {\bibfnamefont {E.~G.}\ \bibnamefont
  {Larsson}}, \bibinfo {author} {\bibfnamefont {O.}~\bibnamefont {Edfors}},
  \bibinfo {author} {\bibfnamefont {F.}~\bibnamefont {Tufvesson}},\ and\
  \bibinfo {author} {\bibfnamefont {T.~L.}\ \bibnamefont {Marzetta}},\
  }\bibfield  {title} {\bibinfo {title} {Massive mimo for next generation
  wireless systems},\ }\href@noop {} {\bibfield  {journal} {\bibinfo  {journal}
  {IEEE communications magazine}\ }\textbf {\bibinfo {volume} {52}},\ \bibinfo
  {pages} {186} (\bibinfo {year} {2014})}\BibitemShut {NoStop}%
\bibitem [{\citenamefont {Nikitopoulos}\ \emph {et~al.}(2014)\citenamefont
  {Nikitopoulos}, \citenamefont {Zhou}, \citenamefont {Congdon},\ and\
  \citenamefont {Jamieson}}]{nikitopoulos2014geosphere}%
  \BibitemOpen
  \bibfield  {author} {\bibinfo {author} {\bibfnamefont {K.}~\bibnamefont
  {Nikitopoulos}}, \bibinfo {author} {\bibfnamefont {J.}~\bibnamefont {Zhou}},
  \bibinfo {author} {\bibfnamefont {B.}~\bibnamefont {Congdon}},\ and\ \bibinfo
  {author} {\bibfnamefont {K.}~\bibnamefont {Jamieson}},\ }\bibfield  {title}
  {\bibinfo {title} {Geosphere: Consistently turning mimo capacity into
  throughput},\ }\href@noop {} {\bibfield  {journal} {\bibinfo  {journal} {ACM
  SIGCOMM Computer Communication Review}\ }\textbf {\bibinfo {volume} {44}},\
  \bibinfo {pages} {631} (\bibinfo {year} {2014})}\BibitemShut {NoStop}%
\bibitem [{\citenamefont {Kim}\ \emph {et~al.}(2021{\natexlab{a}})\citenamefont
  {Kim}, \citenamefont {Kasi}, \citenamefont {Lott}, \citenamefont
  {Venturelli}, \citenamefont {Kaewell},\ and\ \citenamefont
  {Jamieson}}]{kim2021heuristic}%
  \BibitemOpen
  \bibfield  {author} {\bibinfo {author} {\bibfnamefont {M.}~\bibnamefont
  {Kim}}, \bibinfo {author} {\bibfnamefont {S.}~\bibnamefont {Kasi}}, \bibinfo
  {author} {\bibfnamefont {P.~A.}\ \bibnamefont {Lott}}, \bibinfo {author}
  {\bibfnamefont {D.}~\bibnamefont {Venturelli}}, \bibinfo {author}
  {\bibfnamefont {J.}~\bibnamefont {Kaewell}},\ and\ \bibinfo {author}
  {\bibfnamefont {K.}~\bibnamefont {Jamieson}},\ }\bibfield  {title} {\bibinfo
  {title} {Heuristic quantum optimization for 6g wireless communications},\
  }\href@noop {} {\bibfield  {journal} {\bibinfo  {journal} {IEEE Network}\
  }\textbf {\bibinfo {volume} {35}},\ \bibinfo {pages} {8} (\bibinfo {year}
  {2021}{\natexlab{a}})}\BibitemShut {NoStop}%
\bibitem [{\citenamefont {Kim}\ \emph {et~al.}(2019)\citenamefont {Kim},
  \citenamefont {Venturelli},\ and\ \citenamefont
  {Jamieson}}]{kim2019leveraging}%
  \BibitemOpen
  \bibfield  {author} {\bibinfo {author} {\bibfnamefont {M.}~\bibnamefont
  {Kim}}, \bibinfo {author} {\bibfnamefont {D.}~\bibnamefont {Venturelli}},\
  and\ \bibinfo {author} {\bibfnamefont {K.}~\bibnamefont {Jamieson}},\
  }\bibfield  {title} {\bibinfo {title} {Leveraging quantum annealing for large
  mimo processing in centralized radio access networks},\ }in\ \href@noop {}
  {\emph {\bibinfo {booktitle} {Proceedings of the ACM Special Interest Group
  on Data Communication}}}\ (\bibinfo {year} {2019})\ pp.\ \bibinfo {pages}
  {241--255}\BibitemShut {NoStop}%
\bibitem [{\citenamefont {Tabi}\ \emph {et~al.}(2021)\citenamefont {Tabi},
  \citenamefont {Marosits}, \citenamefont {Kallus}, \citenamefont {Vaderna},
  \citenamefont {G{\'o}dor},\ and\ \citenamefont
  {Zimbor{\'a}s}}]{tabi2021evaluation}%
  \BibitemOpen
  \bibfield  {author} {\bibinfo {author} {\bibfnamefont {Z.~I.}\ \bibnamefont
  {Tabi}}, \bibinfo {author} {\bibfnamefont {{\'A}.}~\bibnamefont {Marosits}},
  \bibinfo {author} {\bibfnamefont {Z.}~\bibnamefont {Kallus}}, \bibinfo
  {author} {\bibfnamefont {P.}~\bibnamefont {Vaderna}}, \bibinfo {author}
  {\bibfnamefont {I.}~\bibnamefont {G{\'o}dor}},\ and\ \bibinfo {author}
  {\bibfnamefont {Z.}~\bibnamefont {Zimbor{\'a}s}},\ }\bibfield  {title}
  {\bibinfo {title} {Evaluation of quantum annealer performance via the massive
  mimo problem},\ }\href@noop {} {\bibfield  {journal} {\bibinfo  {journal}
  {IEEE Access}\ }\textbf {\bibinfo {volume} {9}},\ \bibinfo {pages} {131658}
  (\bibinfo {year} {2021})}\BibitemShut {NoStop}%
\bibitem [{\citenamefont {Djidjev}\ \emph {et~al.}(2018)\citenamefont
  {Djidjev}, \citenamefont {Chapuis}, \citenamefont {Hahn},\ and\ \citenamefont
  {Rizk}}]{djidjev2018efficient}%
  \BibitemOpen
  \bibfield  {author} {\bibinfo {author} {\bibfnamefont {H.~N.}\ \bibnamefont
  {Djidjev}}, \bibinfo {author} {\bibfnamefont {G.}~\bibnamefont {Chapuis}},
  \bibinfo {author} {\bibfnamefont {G.}~\bibnamefont {Hahn}},\ and\ \bibinfo
  {author} {\bibfnamefont {G.}~\bibnamefont {Rizk}},\ }\bibfield  {title}
  {\bibinfo {title} {Efficient combinatorial optimization using quantum
  annealing},\ }\href@noop {} {\bibfield  {journal} {\bibinfo  {journal} {arXiv
  preprint arXiv:1801.08653}\ } (\bibinfo {year} {2018})}\BibitemShut {NoStop}%
\bibitem [{\citenamefont {Crosson}\ and\ \citenamefont
  {Harrow}(2016)}]{crosson2016simulated}%
  \BibitemOpen
  \bibfield  {author} {\bibinfo {author} {\bibfnamefont {E.}~\bibnamefont
  {Crosson}}\ and\ \bibinfo {author} {\bibfnamefont {A.~W.}\ \bibnamefont
  {Harrow}},\ }\bibfield  {title} {\bibinfo {title} {Simulated quantum
  annealing can be exponentially faster than classical simulated annealing},\
  }in\ \href@noop {} {\emph {\bibinfo {booktitle} {2016 IEEE 57th Annual
  Symposium on Foundations of Computer Science (FOCS)}}}\ (\bibinfo
  {organization} {IEEE},\ \bibinfo {year} {2016})\ pp.\ \bibinfo {pages}
  {714--723}\BibitemShut {NoStop}%
\bibitem [{\citenamefont {Albash}\ and\ \citenamefont
  {Lidar}(2018)}]{albash2018demonstration}%
  \BibitemOpen
  \bibfield  {author} {\bibinfo {author} {\bibfnamefont {T.}~\bibnamefont
  {Albash}}\ and\ \bibinfo {author} {\bibfnamefont {D.~A.}\ \bibnamefont
  {Lidar}},\ }\bibfield  {title} {\bibinfo {title} {Demonstration of a scaling
  advantage for a quantum annealer over simulated annealing},\ }\href@noop {}
  {\bibfield  {journal} {\bibinfo  {journal} {Physical Review X}\ }\textbf
  {\bibinfo {volume} {8}},\ \bibinfo {pages} {031016} (\bibinfo {year}
  {2018})}\BibitemShut {NoStop}%
\bibitem [{\citenamefont {Kim}\ \emph {et~al.}(2020)\citenamefont {Kim},
  \citenamefont {Venturelli},\ and\ \citenamefont {Jamieson}}]{kim2020towards}%
  \BibitemOpen
  \bibfield  {author} {\bibinfo {author} {\bibfnamefont {M.}~\bibnamefont
  {Kim}}, \bibinfo {author} {\bibfnamefont {D.}~\bibnamefont {Venturelli}},\
  and\ \bibinfo {author} {\bibfnamefont {K.}~\bibnamefont {Jamieson}},\
  }\bibfield  {title} {\bibinfo {title} {Towards hybrid classical-quantum
  computation structures in wirelessly-networked systems},\ }in\ \href@noop {}
  {\emph {\bibinfo {booktitle} {Proceedings of the 19th ACM Workshop on Hot
  Topics in Networks}}}\ (\bibinfo {year} {2020})\ pp.\ \bibinfo {pages}
  {110--116}\BibitemShut {NoStop}%
\bibitem [{\citenamefont {Hegade}\ \emph {et~al.}(2021)\citenamefont {Hegade},
  \citenamefont {Paul}, \citenamefont {Ding}, \citenamefont {Sanz},
  \citenamefont {Albarr{\'a}n-Arriagada}, \citenamefont {Solano},\ and\
  \citenamefont {Chen}}]{hegade2021shortcuts}%
  \BibitemOpen
  \bibfield  {author} {\bibinfo {author} {\bibfnamefont {N.~N.}\ \bibnamefont
  {Hegade}}, \bibinfo {author} {\bibfnamefont {K.}~\bibnamefont {Paul}},
  \bibinfo {author} {\bibfnamefont {Y.}~\bibnamefont {Ding}}, \bibinfo {author}
  {\bibfnamefont {M.}~\bibnamefont {Sanz}}, \bibinfo {author} {\bibfnamefont
  {F.}~\bibnamefont {Albarr{\'a}n-Arriagada}}, \bibinfo {author} {\bibfnamefont
  {E.}~\bibnamefont {Solano}},\ and\ \bibinfo {author} {\bibfnamefont
  {X.}~\bibnamefont {Chen}},\ }\bibfield  {title} {\bibinfo {title} {Shortcuts
  to adiabaticity in digitized adiabatic quantum computing},\ }\href@noop {}
  {\bibfield  {journal} {\bibinfo  {journal} {Physical Review Applied}\
  }\textbf {\bibinfo {volume} {15}},\ \bibinfo {pages} {024038} (\bibinfo
  {year} {2021})}\BibitemShut {NoStop}%
\bibitem [{\citenamefont {King}\ \emph {et~al.}(2018)\citenamefont {King},
  \citenamefont {Bernoudy}, \citenamefont {King}, \citenamefont {Berkley},\
  and\ \citenamefont {Lanting}}]{king2018emulating}%
  \BibitemOpen
  \bibfield  {author} {\bibinfo {author} {\bibfnamefont {A.~D.}\ \bibnamefont
  {King}}, \bibinfo {author} {\bibfnamefont {W.}~\bibnamefont {Bernoudy}},
  \bibinfo {author} {\bibfnamefont {J.}~\bibnamefont {King}}, \bibinfo {author}
  {\bibfnamefont {A.~J.}\ \bibnamefont {Berkley}},\ and\ \bibinfo {author}
  {\bibfnamefont {T.}~\bibnamefont {Lanting}},\ }\bibfield  {title} {\bibinfo
  {title} {Emulating the coherent ising machine with a mean-field algorithm},\
  }\href@noop {} {\bibfield  {journal} {\bibinfo  {journal} {arXiv preprint
  arXiv:1806.08422}\ } (\bibinfo {year} {2018})}\BibitemShut {NoStop}%
\bibitem [{\citenamefont {Marandi}\ \emph {et~al.}(2014)\citenamefont
  {Marandi}, \citenamefont {Wang}, \citenamefont {Takata}, \citenamefont
  {Byer},\ and\ \citenamefont {Yamamoto}}]{marandi2014network}%
  \BibitemOpen
  \bibfield  {author} {\bibinfo {author} {\bibfnamefont {A.}~\bibnamefont
  {Marandi}}, \bibinfo {author} {\bibfnamefont {Z.}~\bibnamefont {Wang}},
  \bibinfo {author} {\bibfnamefont {K.}~\bibnamefont {Takata}}, \bibinfo
  {author} {\bibfnamefont {R.~L.}\ \bibnamefont {Byer}},\ and\ \bibinfo
  {author} {\bibfnamefont {Y.}~\bibnamefont {Yamamoto}},\ }\bibfield  {title}
  {\bibinfo {title} {Network of time-multiplexed optical parametric oscillators
  as a coherent ising machine},\ }\href@noop {} {\bibfield  {journal} {\bibinfo
   {journal} {Nature Photonics}\ }\textbf {\bibinfo {volume} {8}},\ \bibinfo
  {pages} {937} (\bibinfo {year} {2014})}\BibitemShut {NoStop}%
\bibitem [{\citenamefont {Tiunov}\ \emph {et~al.}(2019)\citenamefont {Tiunov},
  \citenamefont {Ulanov},\ and\ \citenamefont {Lvovsky}}]{tiunov2019annealing}%
  \BibitemOpen
  \bibfield  {author} {\bibinfo {author} {\bibfnamefont {E.~S.}\ \bibnamefont
  {Tiunov}}, \bibinfo {author} {\bibfnamefont {A.~E.}\ \bibnamefont {Ulanov}},\
  and\ \bibinfo {author} {\bibfnamefont {A.}~\bibnamefont {Lvovsky}},\
  }\bibfield  {title} {\bibinfo {title} {Annealing by simulating the coherent
  ising machine},\ }\href@noop {} {\bibfield  {journal} {\bibinfo  {journal}
  {Optics express}\ }\textbf {\bibinfo {volume} {27}},\ \bibinfo {pages}
  {10288} (\bibinfo {year} {2019})}\BibitemShut {NoStop}%
\bibitem [{\citenamefont {Reifenstein}\ \emph {et~al.}(2021)\citenamefont
  {Reifenstein}, \citenamefont {Kako}, \citenamefont {Khoyratee}, \citenamefont
  {Leleu},\ and\ \citenamefont {Yamamoto}}]{reifenstein2021coherent}%
  \BibitemOpen
  \bibfield  {author} {\bibinfo {author} {\bibfnamefont {S.}~\bibnamefont
  {Reifenstein}}, \bibinfo {author} {\bibfnamefont {S.}~\bibnamefont {Kako}},
  \bibinfo {author} {\bibfnamefont {F.}~\bibnamefont {Khoyratee}}, \bibinfo
  {author} {\bibfnamefont {T.}~\bibnamefont {Leleu}},\ and\ \bibinfo {author}
  {\bibfnamefont {Y.}~\bibnamefont {Yamamoto}},\ }\bibfield  {title} {\bibinfo
  {title} {Coherent ising machines with optical error correction circuits},\
  }\href@noop {} {\bibfield  {journal} {\bibinfo  {journal} {Advanced Quantum
  Technologies}\ }\textbf {\bibinfo {volume} {4}},\ \bibinfo {pages} {2100077}
  (\bibinfo {year} {2021})}\BibitemShut {NoStop}%
\bibitem [{\citenamefont {Goto}\ \emph {et~al.}(2021)\citenamefont {Goto},
  \citenamefont {Endo}, \citenamefont {Suzuki}, \citenamefont {Sakai},
  \citenamefont {Kanao}, \citenamefont {Hamakawa}, \citenamefont {Hidaka},
  \citenamefont {Yamasaki},\ and\ \citenamefont {Tatsumura}}]{goto2021high}%
  \BibitemOpen
  \bibfield  {author} {\bibinfo {author} {\bibfnamefont {H.}~\bibnamefont
  {Goto}}, \bibinfo {author} {\bibfnamefont {K.}~\bibnamefont {Endo}}, \bibinfo
  {author} {\bibfnamefont {M.}~\bibnamefont {Suzuki}}, \bibinfo {author}
  {\bibfnamefont {Y.}~\bibnamefont {Sakai}}, \bibinfo {author} {\bibfnamefont
  {T.}~\bibnamefont {Kanao}}, \bibinfo {author} {\bibfnamefont
  {Y.}~\bibnamefont {Hamakawa}}, \bibinfo {author} {\bibfnamefont
  {R.}~\bibnamefont {Hidaka}}, \bibinfo {author} {\bibfnamefont
  {M.}~\bibnamefont {Yamasaki}},\ and\ \bibinfo {author} {\bibfnamefont
  {K.}~\bibnamefont {Tatsumura}},\ }\bibfield  {title} {\bibinfo {title}
  {High-performance combinatorial optimization based on classical mechanics},\
  }\href@noop {} {\bibfield  {journal} {\bibinfo  {journal} {Science Advances}\
  }\textbf {\bibinfo {volume} {7}},\ \bibinfo {pages} {eabe7953} (\bibinfo
  {year} {2021})}\BibitemShut {NoStop}%
\bibitem [{\citenamefont {McMahon}\ \emph {et~al.}(2016)\citenamefont
  {McMahon}, \citenamefont {Marandi}, \citenamefont {Haribara}, \citenamefont
  {Hamerly}, \citenamefont {Langrock}, \citenamefont {Tamate}, \citenamefont
  {Inagaki}, \citenamefont {Takesue}, \citenamefont {Utsunomiya}, \citenamefont
  {Aihara} \emph {et~al.}}]{mcmahon2016fully}%
  \BibitemOpen
  \bibfield  {author} {\bibinfo {author} {\bibfnamefont {P.~L.}\ \bibnamefont
  {McMahon}}, \bibinfo {author} {\bibfnamefont {A.}~\bibnamefont {Marandi}},
  \bibinfo {author} {\bibfnamefont {Y.}~\bibnamefont {Haribara}}, \bibinfo
  {author} {\bibfnamefont {R.}~\bibnamefont {Hamerly}}, \bibinfo {author}
  {\bibfnamefont {C.}~\bibnamefont {Langrock}}, \bibinfo {author}
  {\bibfnamefont {S.}~\bibnamefont {Tamate}}, \bibinfo {author} {\bibfnamefont
  {T.}~\bibnamefont {Inagaki}}, \bibinfo {author} {\bibfnamefont
  {H.}~\bibnamefont {Takesue}}, \bibinfo {author} {\bibfnamefont
  {S.}~\bibnamefont {Utsunomiya}}, \bibinfo {author} {\bibfnamefont
  {K.}~\bibnamefont {Aihara}}, \emph {et~al.},\ }\bibfield  {title} {\bibinfo
  {title} {A fully programmable 100-spin coherent ising machine with all-to-all
  connections},\ }\href@noop {} {\bibfield  {journal} {\bibinfo  {journal}
  {Science}\ }\textbf {\bibinfo {volume} {354}},\ \bibinfo {pages} {614}
  (\bibinfo {year} {2016})}\BibitemShut {NoStop}%
\bibitem [{\citenamefont {Kim}\ \emph {et~al.}(2021{\natexlab{b}})\citenamefont
  {Kim}, \citenamefont {Mandr\`{a}}, \citenamefont {Venturelli},\ and\
  \citenamefont {Jamieson}}]{10.1145/3447993.3448619}%
  \BibitemOpen
  \bibfield  {author} {\bibinfo {author} {\bibfnamefont {M.}~\bibnamefont
  {Kim}}, \bibinfo {author} {\bibfnamefont {S.}~\bibnamefont {Mandr\`{a}}},
  \bibinfo {author} {\bibfnamefont {D.}~\bibnamefont {Venturelli}},\ and\
  \bibinfo {author} {\bibfnamefont {K.}~\bibnamefont {Jamieson}},\ }\bibfield
  {title} {\bibinfo {title} {Physics-inspired heuristics for soft mimo
  detection in 5{G} new radio and beyond},\ }in\ \href
  {https://doi.org/10.1145/3447993.3448619} {\emph {\bibinfo {booktitle}
  {Proceedings of the 27th Annual International Conference on Mobile Computing
  and Networking}}},\ \bibinfo {series and number} {MobiCom '21}\ (\bibinfo
  {publisher} {Association for Computing Machinery},\ \bibinfo {address} {New
  York, NY, USA},\ \bibinfo {year} {2021})\ p.\ \bibinfo {pages}
  {42–55}\BibitemShut {NoStop}%
\bibitem [{\citenamefont {Singh}\ \emph {et~al.}(2022)\citenamefont {Singh},
  \citenamefont {Jamieson}, \citenamefont {McMahon},\ and\ \citenamefont
  {Venturelli}}]{singh2022ising}%
  \BibitemOpen
  \bibfield  {author} {\bibinfo {author} {\bibfnamefont {A.~K.}\ \bibnamefont
  {Singh}}, \bibinfo {author} {\bibfnamefont {K.}~\bibnamefont {Jamieson}},
  \bibinfo {author} {\bibfnamefont {P.~L.}\ \bibnamefont {McMahon}},\ and\
  \bibinfo {author} {\bibfnamefont {D.}~\bibnamefont {Venturelli}},\ }\bibfield
   {title} {\bibinfo {title} {Ising machines’ dynamics and regularization for
  near-optimal mimo detection},\ }\href@noop {} {\bibfield  {journal} {\bibinfo
   {journal} {IEEE Transactions on Wireless Communications}\ }\textbf {\bibinfo
  {volume} {21}},\ \bibinfo {pages} {11080} (\bibinfo {year}
  {2022})}\BibitemShut {NoStop}%
\bibitem [{\citenamefont {Norimoto}\ \emph {et~al.}(2023)\citenamefont
  {Norimoto}, \citenamefont {Mori},\ and\ \citenamefont
  {Ishikawa}}]{norimoto2023quantum}%
  \BibitemOpen
  \bibfield  {author} {\bibinfo {author} {\bibfnamefont {M.}~\bibnamefont
  {Norimoto}}, \bibinfo {author} {\bibfnamefont {R.}~\bibnamefont {Mori}},\
  and\ \bibinfo {author} {\bibfnamefont {N.}~\bibnamefont {Ishikawa}},\
  }\bibfield  {title} {\bibinfo {title} {Quantum algorithm for higher-order
  unconstrained binary optimization and mimo maximum likelihood detection},\
  }\href@noop {} {\bibfield  {journal} {\bibinfo  {journal} {IEEE Transactions
  on Communications}\ }\textbf {\bibinfo {volume} {71}},\ \bibinfo {pages}
  {1926} (\bibinfo {year} {2023})}\BibitemShut {NoStop}%
\bibitem [{\citenamefont {Gilliam}\ \emph {et~al.}(2021)\citenamefont
  {Gilliam}, \citenamefont {Woerner},\ and\ \citenamefont
  {Gonciulea}}]{gilliam2021grover}%
  \BibitemOpen
  \bibfield  {author} {\bibinfo {author} {\bibfnamefont {A.}~\bibnamefont
  {Gilliam}}, \bibinfo {author} {\bibfnamefont {S.}~\bibnamefont {Woerner}},\
  and\ \bibinfo {author} {\bibfnamefont {C.}~\bibnamefont {Gonciulea}},\
  }\bibfield  {title} {\bibinfo {title} {Grover adaptive search for constrained
  polynomial binary optimization},\ }\href@noop {} {\bibfield  {journal}
  {\bibinfo  {journal} {Quantum}\ }\textbf {\bibinfo {volume} {5}},\ \bibinfo
  {pages} {428} (\bibinfo {year} {2021})}\BibitemShut {NoStop}%
\bibitem [{\citenamefont {Stoudenmire}\ and\ \citenamefont
  {Waintal}(2023)}]{stoudenmire2023grover}%
  \BibitemOpen
  \bibfield  {author} {\bibinfo {author} {\bibfnamefont {E.}~\bibnamefont
  {Stoudenmire}}\ and\ \bibinfo {author} {\bibfnamefont {X.}~\bibnamefont
  {Waintal}},\ }\bibfield  {title} {\bibinfo {title} {Grover's algorithm offers
  no quantum advantage},\ }\href@noop {} {\bibfield  {journal} {\bibinfo
  {journal} {arXiv preprint arXiv:2303.11317}\ } (\bibinfo {year}
  {2023})}\BibitemShut {NoStop}%
\bibitem [{\citenamefont {Rodríguez-Heck}(2019)}]{rodriguez2018linear}%
  \BibitemOpen
  \bibfield  {author} {\bibinfo {author} {\bibfnamefont {E.}~\bibnamefont
  {Rodríguez-Heck}},\ }\bibfield  {title} {\bibinfo {title} {Linear and
  quadratic reformulations of nonlinear optimization problems in binary
  variables},\ }\href {https://doi.org/10.1007/s10288-018-0392-4} {\bibfield
  {journal} {\bibinfo  {journal} {4OR-A Quarterly Journal of Operations
  Research}\ }\textbf {\bibinfo {volume} {17}},\ \bibinfo {pages} {221}
  (\bibinfo {year} {2019})}\BibitemShut {NoStop}%
\bibitem [{\citenamefont {Mandal}\ \emph {et~al.}(2020)\citenamefont {Mandal},
  \citenamefont {Roy}, \citenamefont {Upadhyay},\ and\ \citenamefont
  {Ushijima-Mwesigwa}}]{mandal2020compressed}%
  \BibitemOpen
  \bibfield  {author} {\bibinfo {author} {\bibfnamefont {A.}~\bibnamefont
  {Mandal}}, \bibinfo {author} {\bibfnamefont {A.}~\bibnamefont {Roy}},
  \bibinfo {author} {\bibfnamefont {S.}~\bibnamefont {Upadhyay}},\ and\
  \bibinfo {author} {\bibfnamefont {H.}~\bibnamefont {Ushijima-Mwesigwa}},\
  }\bibfield  {title} {\bibinfo {title} {Compressed quadratization of higher
  order binary optimization problems},\ }in\ \href@noop {} {\emph {\bibinfo
  {booktitle} {Proceedings of the 17th ACM International Conference on
  Computing Frontiers}}}\ (\bibinfo {year} {2020})\ pp.\ \bibinfo {pages}
  {126--131}\BibitemShut {NoStop}%
\bibitem [{\citenamefont {Verma}\ and\ \citenamefont
  {Lewis}(2022)}]{verma2022penalty}%
  \BibitemOpen
  \bibfield  {author} {\bibinfo {author} {\bibfnamefont {A.}~\bibnamefont
  {Verma}}\ and\ \bibinfo {author} {\bibfnamefont {M.}~\bibnamefont {Lewis}},\
  }\bibfield  {title} {\bibinfo {title} {Penalty and partitioning techniques to
  improve performance of qubo solvers},\ }\href@noop {} {\bibfield  {journal}
  {\bibinfo  {journal} {Discrete Optimization}\ }\textbf {\bibinfo {volume}
  {44}},\ \bibinfo {pages} {100594} (\bibinfo {year} {2022})}\BibitemShut
  {NoStop}%
\bibitem [{\citenamefont {Ayodele}(2022)}]{ayodele2022penalty}%
  \BibitemOpen
  \bibfield  {author} {\bibinfo {author} {\bibfnamefont {M.}~\bibnamefont
  {Ayodele}},\ }\bibfield  {title} {\bibinfo {title} {Penalty weights in qubo
  formulations: Permutation problems},\ }in\ \href@noop {} {\emph {\bibinfo
  {booktitle} {European Conference on Evolutionary Computation in Combinatorial
  Optimization (Part of EvoStar)}}}\ (\bibinfo {organization} {Springer},\
  \bibinfo {year} {2022})\ pp.\ \bibinfo {pages} {159--174}\BibitemShut
  {NoStop}%
\bibitem [{\citenamefont {Garc{\'\i}a}\ \emph {et~al.}(2022)\citenamefont
  {Garc{\'\i}a}, \citenamefont {Ayodele},\ and\ \citenamefont
  {Moraglio}}]{garcia2022exact}%
  \BibitemOpen
  \bibfield  {author} {\bibinfo {author} {\bibfnamefont {M.~D.}\ \bibnamefont
  {Garc{\'\i}a}}, \bibinfo {author} {\bibfnamefont {M.}~\bibnamefont
  {Ayodele}},\ and\ \bibinfo {author} {\bibfnamefont {A.}~\bibnamefont
  {Moraglio}},\ }\bibfield  {title} {\bibinfo {title} {Exact and sequential
  penalty weights in quadratic unconstrained binary optimisation with a digital
  annealer},\ }in\ \href@noop {} {\emph {\bibinfo {booktitle} {Proceedings of
  the Genetic and Evolutionary Computation Conference Companion}}}\ (\bibinfo
  {year} {2022})\ pp.\ \bibinfo {pages} {184--187}\BibitemShut {NoStop}%
\bibitem [{\citenamefont {Damen}\ \emph {et~al.}(2003)\citenamefont {Damen},
  \citenamefont {El~Gamal},\ and\ \citenamefont {Caire}}]{damen2003maximum}%
  \BibitemOpen
  \bibfield  {author} {\bibinfo {author} {\bibfnamefont {M.~O.}\ \bibnamefont
  {Damen}}, \bibinfo {author} {\bibfnamefont {H.}~\bibnamefont {El~Gamal}},\
  and\ \bibinfo {author} {\bibfnamefont {G.}~\bibnamefont {Caire}},\ }\bibfield
   {title} {\bibinfo {title} {On maximum-likelihood detection and the search
  for the closest lattice point},\ }\href@noop {} {\bibfield  {journal}
  {\bibinfo  {journal} {IEEE Transactions on information theory}\ }\textbf
  {\bibinfo {volume} {49}},\ \bibinfo {pages} {2389} (\bibinfo {year}
  {2003})}\BibitemShut {NoStop}%
\bibitem [{\citenamefont {Hammer}\ \emph {et~al.}(1963)\citenamefont {Hammer},
  \citenamefont {Rosenberg}, \citenamefont {Rudeanu} \emph
  {et~al.}}]{hammer1963determination}%
  \BibitemOpen
  \bibfield  {author} {\bibinfo {author} {\bibfnamefont {P.~L.}\ \bibnamefont
  {Hammer}}, \bibinfo {author} {\bibfnamefont {I.}~\bibnamefont {Rosenberg}},
  \bibinfo {author} {\bibfnamefont {S.}~\bibnamefont {Rudeanu}}, \emph
  {et~al.},\ }\bibfield  {title} {\bibinfo {title} {On the determination of the
  minima of pseudo-boolean functions},\ }\href@noop {} {\bibfield  {journal}
  {\bibinfo  {journal} {Studii si Cercetari matematice}\ }\textbf {\bibinfo
  {volume} {14}},\ \bibinfo {pages} {359} (\bibinfo {year} {1963})}\BibitemShut
  {NoStop}%
\bibitem [{\citenamefont {Boros}\ and\ \citenamefont
  {Hammer}(2002)}]{boros2002pseudo}%
  \BibitemOpen
  \bibfield  {author} {\bibinfo {author} {\bibfnamefont {E.}~\bibnamefont
  {Boros}}\ and\ \bibinfo {author} {\bibfnamefont {P.~L.}\ \bibnamefont
  {Hammer}},\ }\bibfield  {title} {\bibinfo {title} {Pseudo-boolean
  optimization},\ }\href@noop {} {\bibfield  {journal} {\bibinfo  {journal}
  {Discrete applied mathematics}\ }\textbf {\bibinfo {volume} {123}},\ \bibinfo
  {pages} {155} (\bibinfo {year} {2002})}\BibitemShut {NoStop}%
\bibitem [{\citenamefont {3GPP}(2018)}]{3gpp2018}%
  \BibitemOpen
  \bibfield  {author} {\bibinfo {author} {\bibnamefont {3GPP}},\ }\href
  {https://www.etsi.org/standards-search} {\emph {\bibinfo {title} {5{G}; {NR};
  Physical Channels and Modulation.}}},\ \bibinfo {type} {Tech. Rep.}\ \bibinfo
  {number} {38.211}\ (\bibinfo  {institution} {3rd Generation Partnership
  Project (3GPP)},\ \bibinfo {year} {2018})\BibitemShut {NoStop}%
\bibitem [{\citenamefont {Borders}\ \emph {et~al.}(2019)\citenamefont
  {Borders}, \citenamefont {Pervaiz}, \citenamefont {Fukami}, \citenamefont
  {Camsari}, \citenamefont {Ohno},\ and\ \citenamefont
  {Datta}}]{borders2019integer}%
  \BibitemOpen
  \bibfield  {author} {\bibinfo {author} {\bibfnamefont {W.~A.}\ \bibnamefont
  {Borders}}, \bibinfo {author} {\bibfnamefont {A.~Z.}\ \bibnamefont
  {Pervaiz}}, \bibinfo {author} {\bibfnamefont {S.}~\bibnamefont {Fukami}},
  \bibinfo {author} {\bibfnamefont {K.~Y.}\ \bibnamefont {Camsari}}, \bibinfo
  {author} {\bibfnamefont {H.}~\bibnamefont {Ohno}},\ and\ \bibinfo {author}
  {\bibfnamefont {S.}~\bibnamefont {Datta}},\ }\bibfield  {title} {\bibinfo
  {title} {Integer factorization using stochastic magnetic tunnel junctions},\
  }\href@noop {} {\bibfield  {journal} {\bibinfo  {journal} {Nature}\ }\textbf
  {\bibinfo {volume} {573}},\ \bibinfo {pages} {390} (\bibinfo {year}
  {2019})}\BibitemShut {NoStop}%
\bibitem [{\citenamefont {Kaiser}\ and\ \citenamefont
  {Datta}(2021)}]{kaiser2021probabilistic}%
  \BibitemOpen
  \bibfield  {author} {\bibinfo {author} {\bibfnamefont {J.}~\bibnamefont
  {Kaiser}}\ and\ \bibinfo {author} {\bibfnamefont {S.}~\bibnamefont {Datta}},\
  }\bibfield  {title} {\bibinfo {title} {Probabilistic computing with p-bits},\
  }\href@noop {} {\bibfield  {journal} {\bibinfo  {journal} {Applied Physics
  Letters}\ }\textbf {\bibinfo {volume} {119}},\ \bibinfo {pages} {150503}
  (\bibinfo {year} {2021})}\BibitemShut {NoStop}%
\bibitem [{\citenamefont {Nikhar}\ \emph {et~al.}(2023)\citenamefont {Nikhar},
  \citenamefont {Kannan}, \citenamefont {Anjum~Aadit}, \citenamefont
  {Chowdhury},\ and\ \citenamefont {Camsari}}]{nikhar2023all}%
  \BibitemOpen
  \bibfield  {author} {\bibinfo {author} {\bibfnamefont {S.}~\bibnamefont
  {Nikhar}}, \bibinfo {author} {\bibfnamefont {S.}~\bibnamefont {Kannan}},
  \bibinfo {author} {\bibfnamefont {N.}~\bibnamefont {Anjum~Aadit}}, \bibinfo
  {author} {\bibfnamefont {S.}~\bibnamefont {Chowdhury}},\ and\ \bibinfo
  {author} {\bibfnamefont {K.~Y.}\ \bibnamefont {Camsari}},\ }\bibfield
  {title} {\bibinfo {title} {All-to-all reconfigurability with sparse and
  higher-order ising machines},\ }\href@noop {} {\bibfield  {journal} {\bibinfo
   {journal} {arXiv e-prints}\ ,\ \bibinfo {pages} {arXiv}} (\bibinfo {year}
  {2023})}\BibitemShut {NoStop}%
\bibitem [{\citenamefont {Onizawa}\ and\ \citenamefont
  {Hanyu}(2024)}]{onizawa2024enhanced}%
  \BibitemOpen
  \bibfield  {author} {\bibinfo {author} {\bibfnamefont {N.}~\bibnamefont
  {Onizawa}}\ and\ \bibinfo {author} {\bibfnamefont {T.}~\bibnamefont
  {Hanyu}},\ }\bibfield  {title} {\bibinfo {title} {Enhanced convergence in
  p-bit based simulated annealing with partial deactivation for large-scale
  combinatorial optimization problems},\ }\href@noop {} {\bibfield  {journal}
  {\bibinfo  {journal} {Scientific Reports}\ }\textbf {\bibinfo {volume}
  {14}},\ \bibinfo {pages} {1339} (\bibinfo {year} {2024})}\BibitemShut
  {NoStop}%
\bibitem [{\citenamefont {Jeon}\ \emph {et~al.}(2015)\citenamefont {Jeon},
  \citenamefont {Ghods}, \citenamefont {Maleki},\ and\ \citenamefont
  {Studer}}]{jeon2015optimality}%
  \BibitemOpen
  \bibfield  {author} {\bibinfo {author} {\bibfnamefont {C.}~\bibnamefont
  {Jeon}}, \bibinfo {author} {\bibfnamefont {R.}~\bibnamefont {Ghods}},
  \bibinfo {author} {\bibfnamefont {A.}~\bibnamefont {Maleki}},\ and\ \bibinfo
  {author} {\bibfnamefont {C.}~\bibnamefont {Studer}},\ }\bibfield  {title}
  {\bibinfo {title} {Optimality of large mimo detection via approximate message
  passing},\ }in\ \href@noop {} {\emph {\bibinfo {booktitle} {2015 IEEE
  International Symposium on Information Theory (ISIT)}}}\ (\bibinfo
  {organization} {IEEE},\ \bibinfo {year} {2015})\ pp.\ \bibinfo {pages}
  {1227--1231}\BibitemShut {NoStop}%
\bibitem [{\citenamefont {Bitra}\ and\ \citenamefont
  {Ponnusamy}(2022)}]{bitra2022large}%
  \BibitemOpen
  \bibfield  {author} {\bibinfo {author} {\bibfnamefont {H.}~\bibnamefont
  {Bitra}}\ and\ \bibinfo {author} {\bibfnamefont {P.}~\bibnamefont
  {Ponnusamy}},\ }\bibfield  {title} {\bibinfo {title} {Large scale mimo
  analysis using enhanced lama},\ }\href@noop {} {\bibfield  {journal}
  {\bibinfo  {journal} {Wireless Personal Communications}\ }\textbf {\bibinfo
  {volume} {126}},\ \bibinfo {pages} {2469} (\bibinfo {year}
  {2022})}\BibitemShut {NoStop}%
\bibitem [{\citenamefont {Burg}\ \emph {et~al.}(2005)\citenamefont {Burg},
  \citenamefont {Borgmann}, \citenamefont {Wenk}, \citenamefont {Zellweger},
  \citenamefont {Fichtner},\ and\ \citenamefont {Bolcskei}}]{burg2005vlsi}%
  \BibitemOpen
  \bibfield  {author} {\bibinfo {author} {\bibfnamefont {A.}~\bibnamefont
  {Burg}}, \bibinfo {author} {\bibfnamefont {M.}~\bibnamefont {Borgmann}},
  \bibinfo {author} {\bibfnamefont {M.}~\bibnamefont {Wenk}}, \bibinfo {author}
  {\bibfnamefont {M.}~\bibnamefont {Zellweger}}, \bibinfo {author}
  {\bibfnamefont {W.}~\bibnamefont {Fichtner}},\ and\ \bibinfo {author}
  {\bibfnamefont {H.}~\bibnamefont {Bolcskei}},\ }\bibfield  {title} {\bibinfo
  {title} {Vlsi implementation of mimo detection using the sphere decoding
  algorithm},\ }\href@noop {} {\bibfield  {journal} {\bibinfo  {journal} {IEEE
  Journal of solid-state circuits}\ }\textbf {\bibinfo {volume} {40}},\
  \bibinfo {pages} {1566} (\bibinfo {year} {2005})}\BibitemShut {NoStop}%
\bibitem [{\citenamefont {He}\ \emph {et~al.}(2023)\citenamefont {He},
  \citenamefont {Kosasih}, \citenamefont {Yu}, \citenamefont {Zhang},
  \citenamefont {Song}, \citenamefont {Hardjawana},\ and\ \citenamefont
  {Letaief}}]{he2023gnn}%
  \BibitemOpen
  \bibfield  {author} {\bibinfo {author} {\bibfnamefont {H.}~\bibnamefont
  {He}}, \bibinfo {author} {\bibfnamefont {A.}~\bibnamefont {Kosasih}},
  \bibinfo {author} {\bibfnamefont {X.}~\bibnamefont {Yu}}, \bibinfo {author}
  {\bibfnamefont {J.}~\bibnamefont {Zhang}}, \bibinfo {author} {\bibfnamefont
  {S.}~\bibnamefont {Song}}, \bibinfo {author} {\bibfnamefont {W.}~\bibnamefont
  {Hardjawana}},\ and\ \bibinfo {author} {\bibfnamefont {K.~B.}\ \bibnamefont
  {Letaief}},\ }\bibfield  {title} {\bibinfo {title} {Gnn-enhanced approximate
  message passing for massive/ultra-massive mimo detection},\ }in\ \href@noop
  {} {\emph {\bibinfo {booktitle} {2023 IEEE Wireless Communications and
  Networking Conference (WCNC)}}}\ (\bibinfo {organization} {IEEE},\ \bibinfo
  {year} {2023})\ pp.\ \bibinfo {pages} {1--6}\BibitemShut {NoStop}%
\bibitem [{\citenamefont {Hung}\ and\ \citenamefont
  {Sang}(2006)}]{hung2006sphere}%
  \BibitemOpen
  \bibfield  {author} {\bibinfo {author} {\bibfnamefont {C.-Y.}\ \bibnamefont
  {Hung}}\ and\ \bibinfo {author} {\bibfnamefont {T.-H.}\ \bibnamefont
  {Sang}},\ }\bibfield  {title} {\bibinfo {title} {A sphere decoding algorithm
  for mimo channels},\ }in\ \href@noop {} {\emph {\bibinfo {booktitle} {2006
  IEEE International Symposium on Signal Processing and Information
  Technology}}}\ (\bibinfo {organization} {IEEE},\ \bibinfo {year} {2006})\
  pp.\ \bibinfo {pages} {502--506}\BibitemShut {NoStop}%
\bibitem [{\citenamefont {Hoydis}\ \emph {et~al.}(2022)\citenamefont {Hoydis},
  \citenamefont {Cammerer}, \citenamefont {Aoudia}, \citenamefont {Vem},
  \citenamefont {Binder}, \citenamefont {Marcus},\ and\ \citenamefont
  {Keller}}]{hoydis2022sionna}%
  \BibitemOpen
  \bibfield  {author} {\bibinfo {author} {\bibfnamefont {J.}~\bibnamefont
  {Hoydis}}, \bibinfo {author} {\bibfnamefont {S.}~\bibnamefont {Cammerer}},
  \bibinfo {author} {\bibfnamefont {F.~A.}\ \bibnamefont {Aoudia}}, \bibinfo
  {author} {\bibfnamefont {A.}~\bibnamefont {Vem}}, \bibinfo {author}
  {\bibfnamefont {N.}~\bibnamefont {Binder}}, \bibinfo {author} {\bibfnamefont
  {G.}~\bibnamefont {Marcus}},\ and\ \bibinfo {author} {\bibfnamefont
  {A.}~\bibnamefont {Keller}},\ }\bibfield  {title} {\bibinfo {title} {Sionna:
  An open-source library for next-generation physical layer research},\
  }\href@noop {} {\bibfield  {journal} {\bibinfo  {journal} {arXiv preprint
  arXiv:2203.11854}\ } (\bibinfo {year} {2022})}\BibitemShut {NoStop}%
\bibitem [{\citenamefont {Swendsen}\ and\ \citenamefont
  {Wang}(1986)}]{swendsen1986replica}%
  \BibitemOpen
  \bibfield  {author} {\bibinfo {author} {\bibfnamefont {R.~H.}\ \bibnamefont
  {Swendsen}}\ and\ \bibinfo {author} {\bibfnamefont {J.-S.}\ \bibnamefont
  {Wang}},\ }\bibfield  {title} {\bibinfo {title} {Replica monte carlo
  simulation of spin-glasses},\ }\href@noop {} {\bibfield  {journal} {\bibinfo
  {journal} {Physical review letters}\ }\textbf {\bibinfo {volume} {57}},\
  \bibinfo {pages} {2607} (\bibinfo {year} {1986})}\BibitemShut {NoStop}%
\bibitem [{\citenamefont {Venturelli}\ and\ \citenamefont
  {Kondratyev}(2019)}]{venturelli2019reverse}%
  \BibitemOpen
  \bibfield  {author} {\bibinfo {author} {\bibfnamefont {D.}~\bibnamefont
  {Venturelli}}\ and\ \bibinfo {author} {\bibfnamefont {A.}~\bibnamefont
  {Kondratyev}},\ }\bibfield  {title} {\bibinfo {title} {Reverse quantum
  annealing approach to portfolio optimization problems},\ }\href@noop {}
  {\bibfield  {journal} {\bibinfo  {journal} {Quantum Machine Intelligence}\
  }\textbf {\bibinfo {volume} {1}},\ \bibinfo {pages} {17} (\bibinfo {year}
  {2019})}\BibitemShut {NoStop}%
\bibitem [{\citenamefont {Beasley}(1998)}]{beasley1998heuristic}%
  \BibitemOpen
  \bibfield  {author} {\bibinfo {author} {\bibfnamefont {J.~E.}\ \bibnamefont
  {Beasley}},\ }\href@noop {} {\emph {\bibinfo {title} {Heuristic algorithms
  for the unconstrained binary quadratic programming problem}}},\ \bibinfo
  {type} {Tech. Rep.}\ (\bibinfo  {institution} {Working Paper, The Management
  School, Imperial College, London, England},\ \bibinfo {year}
  {1998})\BibitemShut {NoStop}%
\bibitem [{\citenamefont {Su}\ \emph {et~al.}(2023)\citenamefont {Su},
  \citenamefont {Kim},\ and\ \citenamefont {Kim}}]{su2023reconfigurable}%
  \BibitemOpen
  \bibfield  {author} {\bibinfo {author} {\bibfnamefont {Y.}~\bibnamefont
  {Su}}, \bibinfo {author} {\bibfnamefont {T.~T.-H.}\ \bibnamefont {Kim}},\
  and\ \bibinfo {author} {\bibfnamefont {B.}~\bibnamefont {Kim}},\ }\bibfield
  {title} {\bibinfo {title} {A reconfigurable cmos ising machine with
  three-body spin interactions for solving boolean satisfiability with direct
  mapping},\ }\href@noop {} {\bibfield  {journal} {\bibinfo  {journal} {IEEE
  Solid-State Circuits Letters}\ } (\bibinfo {year} {2023})}\BibitemShut
  {NoStop}%
\bibitem [{\citenamefont {Bybee}\ \emph {et~al.}(2023)\citenamefont {Bybee},
  \citenamefont {Kleyko}, \citenamefont {Nikonov}, \citenamefont
  {Khosrowshahi}, \citenamefont {Olshausen},\ and\ \citenamefont
  {Sommer}}]{bybee2023efficient}%
  \BibitemOpen
  \bibfield  {author} {\bibinfo {author} {\bibfnamefont {C.}~\bibnamefont
  {Bybee}}, \bibinfo {author} {\bibfnamefont {D.}~\bibnamefont {Kleyko}},
  \bibinfo {author} {\bibfnamefont {D.~E.}\ \bibnamefont {Nikonov}}, \bibinfo
  {author} {\bibfnamefont {A.}~\bibnamefont {Khosrowshahi}}, \bibinfo {author}
  {\bibfnamefont {B.~A.}\ \bibnamefont {Olshausen}},\ and\ \bibinfo {author}
  {\bibfnamefont {F.~T.}\ \bibnamefont {Sommer}},\ }\bibfield  {title}
  {\bibinfo {title} {Efficient optimization with higher-order ising machines},\
  }\href@noop {} {\bibfield  {journal} {\bibinfo  {journal} {Nature
  Communications}\ }\textbf {\bibinfo {volume} {14}},\ \bibinfo {pages} {6033}
  (\bibinfo {year} {2023})}\BibitemShut {NoStop}%
\bibitem [{\citenamefont {Verd{\'u}}(1989)}]{verdu1989computational}%
  \BibitemOpen
  \bibfield  {author} {\bibinfo {author} {\bibfnamefont {S.}~\bibnamefont
  {Verd{\'u}}},\ }\bibfield  {title} {\bibinfo {title} {Computational
  complexity of optimum multiuser detection},\ }\href@noop {} {\bibfield
  {journal} {\bibinfo  {journal} {Algorithmica}\ }\textbf {\bibinfo {volume}
  {4}},\ \bibinfo {pages} {303} (\bibinfo {year} {1989})}\BibitemShut {NoStop}%
\bibitem [{\citenamefont {Gao}\ \emph {et~al.}(2014)\citenamefont {Gao},
  \citenamefont {Lu}, \citenamefont {Han},\ and\ \citenamefont
  {Ning}}]{gao2014near}%
  \BibitemOpen
  \bibfield  {author} {\bibinfo {author} {\bibfnamefont {X.}~\bibnamefont
  {Gao}}, \bibinfo {author} {\bibfnamefont {Z.}~\bibnamefont {Lu}}, \bibinfo
  {author} {\bibfnamefont {Y.}~\bibnamefont {Han}},\ and\ \bibinfo {author}
  {\bibfnamefont {J.}~\bibnamefont {Ning}},\ }\bibfield  {title} {\bibinfo
  {title} {Near-optimal signal detection with low complexity based on
  gauss-seidel method for uplink large-scale mimo systems},\ }in\ \href@noop {}
  {\emph {\bibinfo {booktitle} {2014 IEEE International Symposium on Broadband
  Multimedia Systems and Broadcasting}}}\ (\bibinfo {organization} {IEEE},\
  \bibinfo {year} {2014})\ pp.\ \bibinfo {pages} {1--4}\BibitemShut {NoStop}%
\bibitem [{\citenamefont {Inagaki}\ \emph {et~al.}(2016)\citenamefont
  {Inagaki}, \citenamefont {Haribara}, \citenamefont {Igarashi}, \citenamefont
  {Sonobe}, \citenamefont {Tamate}, \citenamefont {Honjo}, \citenamefont
  {Marandi}, \citenamefont {McMahon}, \citenamefont {Umeki}, \citenamefont
  {Enbutsu} \emph {et~al.}}]{inagaki2016coherent}%
  \BibitemOpen
  \bibfield  {author} {\bibinfo {author} {\bibfnamefont {T.}~\bibnamefont
  {Inagaki}}, \bibinfo {author} {\bibfnamefont {Y.}~\bibnamefont {Haribara}},
  \bibinfo {author} {\bibfnamefont {K.}~\bibnamefont {Igarashi}}, \bibinfo
  {author} {\bibfnamefont {T.}~\bibnamefont {Sonobe}}, \bibinfo {author}
  {\bibfnamefont {S.}~\bibnamefont {Tamate}}, \bibinfo {author} {\bibfnamefont
  {T.}~\bibnamefont {Honjo}}, \bibinfo {author} {\bibfnamefont
  {A.}~\bibnamefont {Marandi}}, \bibinfo {author} {\bibfnamefont {P.~L.}\
  \bibnamefont {McMahon}}, \bibinfo {author} {\bibfnamefont {T.}~\bibnamefont
  {Umeki}}, \bibinfo {author} {\bibfnamefont {K.}~\bibnamefont {Enbutsu}},
  \emph {et~al.},\ }\bibfield  {title} {\bibinfo {title} {A coherent ising
  machine for 2000-node optimization problems},\ }\href@noop {} {\bibfield
  {journal} {\bibinfo  {journal} {Science}\ }\textbf {\bibinfo {volume}
  {354}},\ \bibinfo {pages} {603} (\bibinfo {year} {2016})}\BibitemShut
  {NoStop}%
\bibitem [{\citenamefont {Goto}\ \emph {et~al.}(2019)\citenamefont {Goto},
  \citenamefont {Tatsumura},\ and\ \citenamefont
  {Dixon}}]{goto2019combinatorial}%
  \BibitemOpen
  \bibfield  {author} {\bibinfo {author} {\bibfnamefont {H.}~\bibnamefont
  {Goto}}, \bibinfo {author} {\bibfnamefont {K.}~\bibnamefont {Tatsumura}},\
  and\ \bibinfo {author} {\bibfnamefont {A.~R.}\ \bibnamefont {Dixon}},\
  }\bibfield  {title} {\bibinfo {title} {Combinatorial optimization by
  simulating adiabatic bifurcations in nonlinear hamiltonian systems},\
  }\href@noop {} {\bibfield  {journal} {\bibinfo  {journal} {Science advances}\
  }\textbf {\bibinfo {volume} {5}},\ \bibinfo {pages} {eaav2372} (\bibinfo
  {year} {2019})}\BibitemShut {NoStop}%
\bibitem [{\citenamefont {Bulger}\ \emph {et~al.}(2003)\citenamefont {Bulger},
  \citenamefont {Baritompa},\ and\ \citenamefont
  {Wood}}]{bulger2003implementing}%
  \BibitemOpen
  \bibfield  {author} {\bibinfo {author} {\bibfnamefont {D.}~\bibnamefont
  {Bulger}}, \bibinfo {author} {\bibfnamefont {W.~P.}\ \bibnamefont
  {Baritompa}},\ and\ \bibinfo {author} {\bibfnamefont {G.~R.}\ \bibnamefont
  {Wood}},\ }\bibfield  {title} {\bibinfo {title} {Implementing pure adaptive
  search with grover's quantum algorithm},\ }\href@noop {} {\bibfield
  {journal} {\bibinfo  {journal} {Journal of optimization theory and
  applications}\ }\textbf {\bibinfo {volume} {116}},\ \bibinfo {pages} {517}
  (\bibinfo {year} {2003})}\BibitemShut {NoStop}%
\bibitem [{\citenamefont {Babbush}\ \emph {et~al.}(2021)\citenamefont
  {Babbush}, \citenamefont {McClean}, \citenamefont {Newman}, \citenamefont
  {Gidney}, \citenamefont {Boixo},\ and\ \citenamefont
  {Neven}}]{babbush2021focus}%
  \BibitemOpen
  \bibfield  {author} {\bibinfo {author} {\bibfnamefont {R.}~\bibnamefont
  {Babbush}}, \bibinfo {author} {\bibfnamefont {J.~R.}\ \bibnamefont
  {McClean}}, \bibinfo {author} {\bibfnamefont {M.}~\bibnamefont {Newman}},
  \bibinfo {author} {\bibfnamefont {C.}~\bibnamefont {Gidney}}, \bibinfo
  {author} {\bibfnamefont {S.}~\bibnamefont {Boixo}},\ and\ \bibinfo {author}
  {\bibfnamefont {H.}~\bibnamefont {Neven}},\ }\bibfield  {title} {\bibinfo
  {title} {Focus beyond quadratic speedups for error-corrected quantum
  advantage},\ }\href@noop {} {\bibfield  {journal} {\bibinfo  {journal} {PRX
  quantum}\ }\textbf {\bibinfo {volume} {2}},\ \bibinfo {pages} {010103}
  (\bibinfo {year} {2021})}\BibitemShut {NoStop}%
\bibitem [{\citenamefont {Isakov}\ \emph {et~al.}(2015)\citenamefont {Isakov},
  \citenamefont {Zintchenko}, \citenamefont {R{\o}nnow},\ and\ \citenamefont
  {Troyer}}]{isakov2015optimised}%
  \BibitemOpen
  \bibfield  {author} {\bibinfo {author} {\bibfnamefont {S.~V.}\ \bibnamefont
  {Isakov}}, \bibinfo {author} {\bibfnamefont {I.~N.}\ \bibnamefont
  {Zintchenko}}, \bibinfo {author} {\bibfnamefont {T.~F.}\ \bibnamefont
  {R{\o}nnow}},\ and\ \bibinfo {author} {\bibfnamefont {M.}~\bibnamefont
  {Troyer}},\ }\bibfield  {title} {\bibinfo {title} {Optimised simulated
  annealing for ising spin glasses},\ }\href@noop {} {\bibfield  {journal}
  {\bibinfo  {journal} {Computer Physics Communications}\ }\textbf {\bibinfo
  {volume} {192}},\ \bibinfo {pages} {265} (\bibinfo {year}
  {2015})}\BibitemShut {NoStop}%
\end{thebibliography}
%
\end{document}